\documentclass[review,preprint]{elsarticle}
\nopreprintlinetrue

\usepackage{hyperref}
\usepackage{amsmath,amssymb,amsfonts}
\usepackage{graphicx}
\usepackage{textcomp}
\usepackage{algorithm}
\usepackage{algpseudocode}
\usepackage[normalem]{ulem}
\usepackage{bm}
\usepackage[table]{xcolor}
\usepackage{tabularx}
\usepackage{booktabs}
\usepackage{siunitx}
\usepackage{array}
\usepackage{caption}
\usepackage{multicol}
\usepackage{xcolor}
\usepackage{ulem}
\usepackage{lineno}

\begin{document}

\begin{frontmatter}



\title{Predicting Diabetic Macular Edema Treatment Responses Using OCT: Dataset and Methods of APTOS Competition}

\author[1]{Weiyi Zhang\textsuperscript{1,2,}}
\author[2]{Peranut Chotcomwongse\textsuperscript{1,2,}}
\author[11]{Yinwen Li\textsuperscript{2,}}
\author[1]{Pusheng Xu\textsuperscript{2,}}

\author[4]{Ruijie Yao}
\author[5]{Lianhao Zhou}
\author[6]{Yuxuan Zhou}
\author[7]{Hui Feng}
\author[8]{Qiping Zhou}

\author[1,15]{Xinyue Wang\textsuperscript{2,}}
\author[16]{Shoujin Huang\textsuperscript{2,}}
\author[12]{Zihao Jin\textsuperscript{2,}}
\author[3]{Florence H.T. Chung\textsuperscript{2,}}

\author[13]{Shujun Wang\textsuperscript{2,}}
\author[14]{Yalin Zheng\textsuperscript{2,}}

\author[1,9,10]{Mingguang He\textsuperscript{2,}\corref{cor1}}
\author[1,9]{Danli Shi\textsuperscript{2,}\corref{cor1}}
\author[2]{Paisan Ruamviboonsuk\textsuperscript{2,}\corref{cor1}}

\address[1]{School of Optometry, The Hong Kong Polytechnic University, Hong Kong}
\address[2]{Department of Ophthalmology, College of Medicine, Rangsit University, Rajavithi Hospital, Thailand}
\address[3]{Asia Pacific Tele-Ophthalmology Society, c/o State Key Laboratory (Ophthalmology), Zhongshan Ophthalmic Center, Sun Yat-Sen University, China}
\address[4]{Duke University, USA}
\address[5]{Texas A\&M University, USA}
\address[6]{Department of Electronic Engineering, Tsinghua University, China}
\address[7]{Xidian University, China}
\address[8]{Svision Imaging Ltd., China}
\address[9]{Research Centre for SHARP Vision (RCSV), The Hong Kong Polytechnic University, Hong Kong}
\address[10]{Centre for Eye and Vision Research (CEVR), 17W Hong Kong Science Park, Hong Kong}
\address[11]{Department of Ophthalmology and Visual Sciences, the Chinese University of Hong Kong, Hong Kong}
\address[12]{Faculty of Engineering, The University of Hong Kong, Hong Kong}
\address[13]{Faculty of Engineering, The Hong Kong Polytechnic University, Hong Kong}
\address[14]{Faculty of Health and Life Sciences, University of Liverpool, United Kingdom}
\address[15]{Chien-Shiung Wu College, Southeast University, China}
\address[16]{Shenzhen Technology University, China}

\fntext[1]{These authors contributed equally to this work.}
\fntext[2]{These authors co-organized the competition. All others contributed methods of their algorithm(s) presented in the paper.} 

\cortext[cor1]{Corresponding author}

\begin{abstract}
Diabetic macular edema (DME) significantly contributes to visual impairment in diabetic patients. Treatment responses to intravitreal therapies vary, highlighting the need for patient stratification to predict therapeutic benefits and enable personalized strategies. To our knowledge, this study is the first to explore pre-treatment stratification for predicting DME treatment responses. To advance this research, we organized the 2nd Asia-Pacific Tele-Ophthalmology Society (APTOS) Big Data Competition in 2021. The competition focused on improving predictive accuracy for anti-VEGF therapy responses using ophthalmic OCT images. We provided a dataset containing tens of thousands of OCT images from 2,000 patients with labels across four sub-tasks. This paper details the competition's structure, dataset, leading methods, and evaluation metrics. The competition attracted strong scientific community participation, with 170 teams initially registering and 41 reaching the final round. The top-performing team achieved an AUC of 80.06\%, highlighting the potential of AI in personalized DME treatment and clinical decision-making.
\end{abstract}




\end{frontmatter}



\section{Introduction}
\label{sec1}
Diabetic retinopathy (DR) is a prevalent complication among individuals with diabetes, affecting approximately one-third of diabetic patients \cite{RN1}. Among patients with DR, diabetes macular edema (DME) stands out as a leading cause of visual impairment, with a growing global incidence affecting over 19 million individuals \cite{RN2}. It is also the primary reason for vision loss among working-age adults worldwide. Effective and timely treatment is essential to preserving visual function and preventing disease progression. At present, intravitreal anti-vascular endothelial growth factor (anti-VEGF) injection has emerged as the standard therapy for treating DME \cite{RN3}. This is attributed to its numerous clinical advantages, which encompass enhancements in both function and structure. Moreover, it has the potential to arrest or even reverse the progression of DR \cite{RN4}. Nonetheless, patient responses vary considerably, with a significant proportion of patients showing suboptimal outcomes. According to the Diabetic Retinopathy Clinical Research (DRCR) Retina Network Protocol T study, it has been observed that 31.6\% to 65.6\% of patients with DME continue to experience persistent edema even after receiving regular intravitreal injections for a minimum of four times over a span of 24 weeks \cite{RN5}.

Recently, optical coherence tomography (OCT) has become the gold standard for diagnosing and monitoring treatment responses in diabetic macular edema (DME) due to its high-resolution imaging and precise quantitative assessments of retinal thickness and microscopic pathological changes \cite{RN6,RN7}. Compared to traditional methods such as slit-lamp biomicroscopy and fundus photography, OCT provides a more detailed evaluation, improving diagnostic accuracy and facilitating personalized therapeutic strategies \cite{RN8,RN9}. However, interpreting OCT scans requires specialized expertise, and the shortage of experienced ophthalmologists often leads to diagnostic inefficiencies. This underscores the need for automated OCT-based prediction systems to enhance patient stratification and optimize treatment decisions.

The integration of artificial intelligence (AI) into ophthalmology has driven advancements in automated diagnosis \cite{transffa,ffagpt,transicga}. While OCT is the preferred imaging modality for DME diagnosis, assessing therapeutic responses remains challenging and often prone to variability. AI-powered automated analysis can alleviate the burden on ophthalmologists by improving diagnostic accuracy, standardizing assessments, and enhancing clinical efficiency \cite{RN14}. Based on this concept, numerous studies have focused on the use of AI for intelligent diagnosis of DME via OCT. Tang et al. proposed a deep learning system for the automated classification of DME based on both three-dimensional (3D) volume scans and 2D B-scan OCT images, yielding volume scan-level results for each eye \cite{RN15}. Midena et al. detected and quantified different OCT biomarkers in the eyes of DME patients, proposing a reliable and reproducible AI algorithm tool \cite{RN16}. Tripathi et al. employed an automatic method based on Generative Adversarial Networks (GANs) to generate OCT B-scan images of DME, to enhance the reliability of the DME detection system \cite{RN17}. 

However, most existing studies focus on the classification of DME rather than predicting treatment outcomes for patients undergoing anti-VEGF therapy. Moreover, the majority of related datasets are private \cite{tang2017private,RN15,RN16}, and publicly available datasets are extremely limited in scale. For instance, the Duke Eye dataset \cite{srinivasan2014small} includes only 15 subjects with DME, making it insufficient for robust deep learning research. A significant challenge in this field is the lack of large, publicly accessible datasets, which not only hinders the development and validation of deep learning models for predicting treatment responses, but also makes it difficult to fairly compare the performance of different approaches. This data scarcity restricts researchers’ ability to design and evaluate algorithms effectively, ultimately limiting the clinical translation and advancement of AI-driven approaches for DME treatment prediction.

To address this gap, we collected and released a large-scale, publicly accessible OCT dataset specifically aimed at facilitating AI research for predicting treatment outcomes in DME patients undergoing anti-VEGF therapy. Additionally, we organized the 2nd Asia-Pacific Tele-Ophthalmology Society (APTOS) Big Data Competition in 2021. This competition aimed to foster advancements in AI-driven diagnostic algorithms for ophthalmic OCT, particularly enhancing predictive accuracy for patient responses to anti-VEGF treatments. The overarching objective of this initiative is to support ophthalmologists in developing individualized treatment plans and promoting timely and effective patient care. In this paper, we introduce the dataset, describe the competition setup, summarize the participating solutions, and present the results and discussion.

\section{Related Work}

In this section, we review previous challenges and associated datasets for retinal image analysis and summarize existing AI methodologies applied to OCT image analysis.

\subsection{Review of Previous Challenges}
Several public challenges have been organized for retinal image analysis (Table \ref{tab1}), with the majority focusing on color fundus photography (CFP). To the best of our knowledge, the Retinopathy Online Challenge (ROC) held in 2009 was the first competition targeting retinal images for automated microaneurysm detection \cite{niemeijer2009retinopathy}. Subsequent initiatives included the Diabetic Retinopathy Detection competition on Kaggle in 2015 \cite{diabetic-retinopathy-detection} and the APTOS 2019 Blindness Detection challenge \cite{kobat2022automated}, both aimed at diabetic retinopathy (DR) severity grading. The Indian Diabetic Retinopathy Image Dataset (IDRiD) challenge \cite{porwal2020idrid} extended the scope to include both DR grading and lesion segmentation, while DeepDRiD further incorporated image quality assessment and transfer learning from CFP to ultra-widefield (UWF) images \cite{liu2022deepdrid}. The DRAC 2022 (Diabetic Retinopathy Analysis Challenge) featured tasks including image quality grading, DR severity classification, and lesion segmentation on ultra-widefield OCTA images \cite{qian2024drac}.

Beyond DR, other challenges addressed various ophthalmic conditions. The REFUGE (Retinal Fundus Glaucoma Challenge) and REFUGE2 targeted glaucoma detection and optic disc/cup segmentation using fundus images \cite{orlando2020refuge, fang2022refuge2}. The AIROGS (Artificial Intelligence for RObust Glaucoma Screening) challenge \cite{de2023airogs} and JustRAIGS (Justified Referral in AI Glaucoma Screening) challenge \cite{lemij2023characteristics} focused on glaucoma detection, with the latter also involving multi-label classification. Multi-disease classification was explored in the Ocular Disease Intelligent Recognition (ODIR-2019) and Retinal Image Analysis for Multi-Disease Detection (RIADD) challenges \cite{ODIR-2019, pachade2025rfmid}. In the context of pathological myopia, the PAthoLogic Myopia (PALM) challenge provided datasets for both disease detection and lesion segmentation \cite{fang2024open}. The ADAM (Automatic Detection challenge on Age-related Macular degeneration) challenge addressed age-related macular degeneration (AMD), including fovea localization and segmentation of optic disc and macular lesions \cite{fang2022adam}. The AGE (Angle-closure Glaucoma Evaluation) challenge focused on angle-closure classification and scleral spur localization in anterior segment OCT (AS-OCT) images \cite{fu2020age}. More recently, the 2023 APTOS Big Data Competition focused on report generation from fundus fluorescein angiography (FFA) images \cite{zhang2023angiographic}.

In addition to CFP-based challenges, a number of competitions have been devoted to OCT images. For instance, the Retinal OCT Fluid Challenge (RETOUCH) \cite{bogunovic2019retouch} addressed the detection and segmentation of intraretinal fluid (IRF), subretinal fluid (SRF), and pigment epithelial detachment (PED) in retinal OCT scans. A total of 112 macula-centered OCT volumes were included, with macular edema attributed to AMD in half of the cases and to retinal vein occlusion (RVO) in the other half. The Retinal OCT Classification Challenge (ROCC) \cite{ROCC-2017} focused on DR detection using OCT, while the GAMMA challenge \cite{wu2023gamma} encompassed glaucoma grading, fovea localization, and optic disc/cup segmentation using both CFP and OCT modalities. Notably, the 2024 APTOS Big Data Competition emphasized OCT image generation from CFP \cite{APTOS-2024}.

\begin{table}[t]
\caption{Previous public challenges held for retinal image analysis. OCT = Optical Coherence Tomography (refers to posterior segment OCT unless otherwise specified); CFP = Color Fundus Photography; AS-OCT = Anterior Segment Optical Coherence Tomography; UWF = ultra-widefield; OCTA = Optical Coherence Tomography Angiography; FFA = Fundus Fluorescein Angiography.}
\centering
\resizebox{\textwidth}{!}{%
\begin{tabular}{p{1.2cm} p{3.8cm} p{1.4cm} p{4.3cm} p{3.5cm} p{4cm}}
\toprule
\textbf{Year} & \textbf{Competition Name} & \textbf{Modality} & \textbf{Task} & \textbf{Dataset} & \textbf{Link} \\
\midrule
2015 & Diabetic Retinopathy Detection (2015) & CFP & DR grading (5-level classification) & 35k images & https://www.kaggle.com/c/diabetic-retinopathy-detection \\
2017 & RETOUCH challenge & OCT & Detection and segmentation of retinal OCT fluid & 112 OCT volumes (11,334 B-scans) & https://retouch.grand-challenge.org/ \\
2017 & Retinal OCT Classification Challenge (ROCC) & OCT & DR detection & 165 images & https://rocc.grand-challenge.org/ \\
2018 & IDRiD – DR Segmentation \& Grading (ISBI 2018) & CFP & Lesion segmentation, grading, disc/fovea localization. & 54 images & https://idrid.grand-challenge.org \\
2018 & REFUGE challenge & CFP & Glaucoma detection and disc/cup segmentation  & 1,200 images & https://refuge.grand-challenge.org \\
2019 & PALM: PAthoLogic Myopia Challenge & CFP & Pathological myopia detection (classification) and lesion segmentation (atrophy, etc.) & 1,200 images & https://palm.grand-challenge.org \\
2019 & AGE challenge & AS-OCT & Angle closure classification and localization of scleral spur & 4,800 images & https://age.grand-challenge.org/ \\
2019 & APTOS 2019 Blindness Detection & CFP & DR grading (5-level classification) & 3,662 images & https://www.kaggle.com/competitions/aptos2019-blindness-detection \\
2019 & Ocular Disease Intelligent Recognition (ODIR-2019) & CFP & Ocular disease recognition & 10,000 images & https://odir2019.grand-challenge.org/ \\
2020 & REFUGE2 challenge & CFP & Glaucoma detection and disc/cup segmentation  & 2,000 images & https://refuge.grand-challenge.org/ \\
2020 & ADAM challenge & CFP & Detection of AMD, localization of fovea, and segmentation of optic disc and lesions & 1,200 images & https://amd.grand-challenge.org/ \\
2020 & ISBI 2020: Diabetic Retinopathy—Grading and Image Quality Estimation Challenge & CFP+UWF & DR grading, image quality estimation and transfer learning (cross-domain DR grading) & 2,000 CFP and 256 UWF images & https://isbi.deepdr.org \\
2021 & RIADD challenge & CFP & Ocular disease screening and classification & 3,200 images & https://riadd.grand-challenge.org/ \\
2021 & GAMMA Challenge & CFP+OCT & Glaucoma grading, fovea localization, and disc/cup segmentation & 300 patients (CFP + OCT) & https://gamma.grand-challenge.org \\
2022 & AIROGS challenge & CFP & Glaucoma detection & 113,893 images & https://airogs.grand-challenge.org/ \\
2022 & DRAC 2022 Diabetic Retinopathy Analysis Challenge & OCTA & Image quality assessment, DR grading and lesion segmentation & 1,103 UWF-OCTA images & https://drac22.grand-challenge.org \\
2023 & 2023 APTOS Big Data Competition & FFA & Report generation from FFA images & 50,000 images & https://2023.asiateleophth.org/big-data-competition/ \\
2024 & JustRAIGS challenge & CFP & Glaucoma detection and multi-label classification & 110k images & https://justraigs.grand-challenge.org/ \\
2024 & 2024 APTOS Big Data Competition & CFP+OCT & Generation of OCT images from CFP & 1,271 pairs of CFP and OCT images from 693 patients & https://2024.asiateleophth.org/big-data-competition/ \\

\bottomrule
\end{tabular}
}
\label{tab1}
\end{table}

\subsection{AI Methods for OCT Image Analysis}
The application of AI for OCT image analysis has progressed from traditional machine learning to deep neural networks\cite{RN34,RN51}, with a more recent focus on large fundamental models\cite{RN50}. Early OCT image analysis used traditional machine learning, manually extracting features(thickness of the retina, volume of the pathologies\cite{RN52}) for algorithms (logistic regression\cite{RN53}, PCA\cite{RN54}, SVMs, Random Forests\cite{RN52}) and use them for disease classification and basic segmentation. However, the performance of these methods was limited by the quality of feature engineering and generalization ability. Subsequently, deep learning with CNNs became dominant. Successful architectures like VGG\cite{RN55,RN56} and ResNet\cite{RN57,RN58} can classify diseases, such as DR and AMD, in OCT images by extracting complex image features directly, often with better performance compared to older techniques. Furthermore, the development and application of specialized architectures such as U-Net and its variants have made great contributions to retinal layer detection and anatomical segmentation\cite{RN59}, providing remarkable performance\cite{RN60} on commonly used datasets.

Recent AI advancements in retinal image analysis focus on large fundamental models, including pre-trained models \cite{shi2024eyefound}, multi-modal models \cite{transicga}, and autoregressive models \cite{zhang2024fundus2video}, to improve generalizability and address data scarcity. Multimodal data integration\cite{RN61} shows promise for better zero-shot capability in long-tail learning scenarios. GPT-4V and chatbot\cite{RN50,RN62,RN63} interpretations also demonstrate significant potential for more comprehensive ophthalmic analysis, extending beyond image interpretation to case report generation and preliminary treatment recommendations, yet concerns regarding clinical applicability and interpretability persist.

\section{Dataset: OCT4DME}

\subsection{Data Overview}
The OCT4DME dataset was utilized in the APTOS 2021 Big Data Competition, which was primarily centered on predicting the outcomes of anti-VEGF treatment in patients with diabetic macular edema (DME). This dataset consists of tens of thousands of OCT images from 2,000 patients, both before and after undergoing initial anti-VEGF treatments. These treatments encompass a group of medications designed to reduce the growth of new blood vessels or swelling and are employed in the treatment of various eye conditions characterized by neovascularization or edema in the macular region of the retina. Authorized users of the APTOS cross-country datasets have the privilege of utilizing these datasets for constructing machine learning models and training their AI algorithms.

\subsection{Data Acquisition}
The data was provided by the Asia Pacific Tele-Ophthalmology Society, Rajavithi Hospital (Thailand), and an anonymous Chinese cohort. Data was collected between January 1, 2015, and December 31, 2020. Inclusion criteria was as follows: (1) patients aged 18 years or older diagnosed with macular neovascularization; (2) patients who received at least one intravitreal anti-vascular endothelial growth factor (anti-VEGF) injection and completed a follow-up visit at least 6 months after the initial injection; and (3) patients who underwent OCT imaging within 3 months prior to the first anti-VEGF injection. Exclusion criteria included: (1) absence of recorded visual acuity; (2) a history of retinal treatment involving laser therapy or intraocular surgery; and (3) presence of coexisting ocular conditions that could affect visual acuity (e.g., advanced glaucoma, diabetic retinopathy, or significant media opacity).

OCT images with a resolution of 768 × 768 pixels were acquired using a 25-line raster scan protocol on the Heidelberg Spectralis OCT2 plus system (Heidelberg Engineering, Germany). In instances where raster scans were unavailable, 6-line radial scans were utilized as an alternative. Collection and annotation for OCT4DME were carried out in two stages. Specifically, during the first stage of the competition, images and labels from 2,366 eyes were employed for training set, while an additional 361 eyes were designated for testing set. In the subsequent stage, training set involved scans from 221 eyes, with the remaining 342 eyes reserved for testing set. OCT4DME is distributed under \href{https://creativecommons.org/licenses/by-nc-sa/4.0/}{CC BY-NC-SA 4.0}.

\subsection{Data Statistics}

\begin{figure}[t]
\centering
\includegraphics[width=0.8\textwidth]{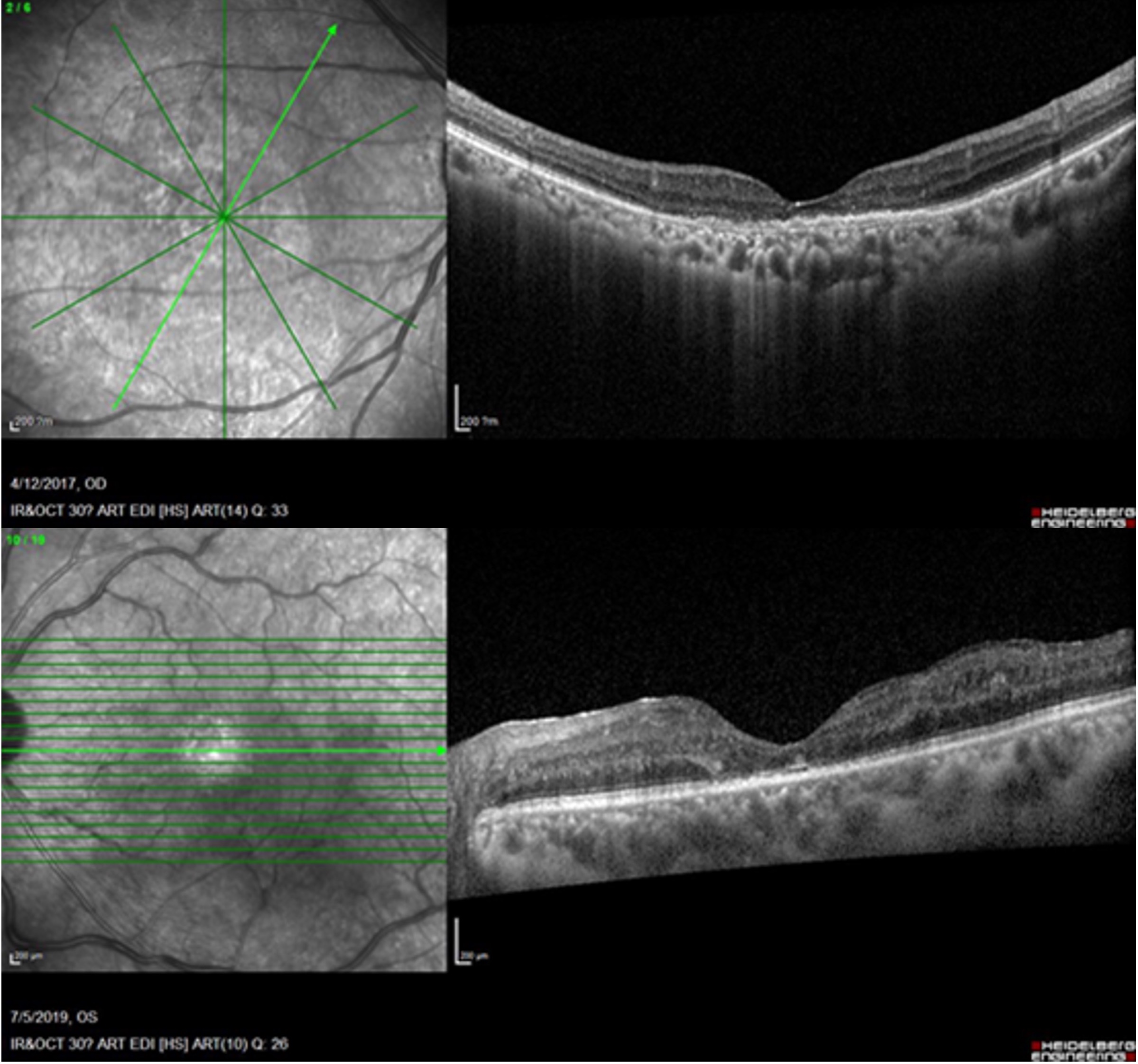}
\caption{An example from the dataset OCT4DME. } \label{fig1}
\end{figure}

Figure \ref{fig1} shows the two parts in each image from OCT4DME. The left is the fundus image, with green (or red) lines indicating the positions where the tomography images are taken. The right is the OCT image, which renders an in vivo cross-sectional view of the retina. The bright line with an arrow in the fundus image shows the position of the scan line for this OCT image. The full number of images (scans) for each eye (pre-treatment or post-treatment) is 6 with radial scan style or 19/25/31 with horizontal scan style. In many cases, 1 - 3 images for an eye from the first stage are missing while only several (2-10) horizontal scans are available for each eye from the subsequent stage.

\subsection{Data Annotation}
Automated central subfield thickness (CST) measurements were obtained using the Heidelberg Spectralis software. All automated segmentations were independently reviewed and confirmed by trained graders. In cases where only six-line raster scans were available rather than volume scans, central retinal thickness (CRT) was manually measured as the perpendicular distance from the internal limiting membrane (ILM) to the outer border of the retinal pigment epithelium (RPE). Image annotations were performed by two retina fellows-in-training under the supervision of a retinal specialist with over 25 years of experience (PR). Annotations were iteratively reviewed and refined until consensus was reached among the graders. For quality assurance, approximately 20\% of the annotated images were randomly selected and audited by PR. The labels for each eye are listed in Table \ref{tab2}. For eyes in the first stage, all the labels are collected in a per-eye style, where Intraretinal Fluid before treatment (preIRF), Subretinal Fluid before treatment (preSRF), Pigment Epithelial Detachment before treatment (prePED), Hyper Reflective Foci before treatment (preHRF), Intraretinal Fluid after treatment (IRF), Subretinal Fluid after treatment (SRF), Pigment Epithelial Detachment after treatment (PED) and Hyper Reflective Foci after treatment (HRF) are annotated once for the whole group of scans. However, in the second stage, IRF, SRF, PED, and HRF are annotated in a per-picture style. Each scan has its own annotations for these 4 items. The completeness of the annotations for the validation set in each of the two stages is guaranteed to facilitate evaluation. Several labels may be missing for some eyes in the training set.


\begin{table}[t]
\caption{Labels for each eye in the dataset OCT4DME.}
\centering
\resizebox{\textwidth}{!}{%
\begin{tabular}{ll}
\hline
\textbf{Label}          & \textbf{Annotation}                                                                                                                                                                                                                                                                                                                \\ \hline
patient ID              & PatientID+EyeSide (L: left/R: right)                                                                                                                                                                                                                                                                                               \\
gender                  & 1 for male or 2 for female                                                                                                                                                                                                                                                                                                         \\
age                     & age of the patient                                                                                                                                                                                                                                                                                                                 \\
diagnosis               & \begin{tabular}[c]{@{}l@{}}1=wet AMD=CNVM, 2=PCV, 3=DME, 4=RVO, 5=CME,   \\ 6=fellow eye (not receive anti-VEGF), 9=other diagnosis\end{tabular}                                                                                                                                                                                   \\
anti-VEGF               & \begin{tabular}[c]{@{}l@{}}1=bevacizumab=Avastin,   2=ranibizumab=Accentrix=Razumab, \\ 3=aflibercept=Eylea, 4=conbercept,   5=triamcinolone=Tricort, \\ 6=combination, 9=other treatment, 0=not receiving   Anti-VEGF, \\ Ozurdex=dexamethasone (new drug in the second stage), \\ Pagenax=brolucizumab (new drug in the second stage)\end{tabular} \\
preVA                   & Visual Acuity (logMAR) before treatment                                                                                                                                                                                                                                                                                            \\
preCST                  & Central Subfield Thickness before treatment                                                                                                                                                                                                                                                                                        \\
preIRF                  & Intraretinal Fluid before treatment                                                                                                                                                                                                                                                                                                \\
preSRF                  & Subretinal Fluid before treatment                                                                                                                                                                                                                                                                                                  \\
prePED                  & Pigment Epithelial Detachment before treatment                                                                                                                                                                                                                                                                                     \\
preHRF                  & Hyper Reflective Foci before treatment                                                                                                                                                                                                                                                                                             \\
VA                      & Visual Acuity (logMAR) after treatment                                                                                                                                                                                                                                                                                             \\
CST                     & Central Subfield Thickness after treatment                                                                                                                                                                                                                                                                                         \\
IRF                     & Intraretinal Fluid after treatment                                                                                                                                                                                                                                                                                                 \\
SRF                     & Subretinal Fluid after treatment                                                                                                                                                                                                                                                                                                   \\
PED                     & Pigment Epithelial Detachment after treatment                                                                                                                                                                                                                                                                                      \\
HRF                     & Hyper Reflective Foci after treatment                                                                                                                                                                                                                                                                                              \\
Continue Injection (CI) & whether anti-VEGF injection is continued or not                                                                                                                                                                                                                                                                                    \\ \hline
\end{tabular}
}
\label{tab2}
\end{table}

\section{Competition Setup}
\subsection{Challenge Organizers}
The 2021 Asia Pacific Tele-Ophthalmology Society (APTOS) Big Data Competition was hosted by the APTOS and co-organized by Rajavithi Hospital in Thailand, Aravind Eye Hospital in India, and Zhongshan Ophthalmic Center at Sun Yat-sen University in China. It was sponsored and recognized by the Medical Services Department of the Ministry of Public Health of Thailand, and was exclusively supported by the Alibaba Cloud Tianchi platform.
\begin{figure}[t]
\includegraphics[width=\textwidth]{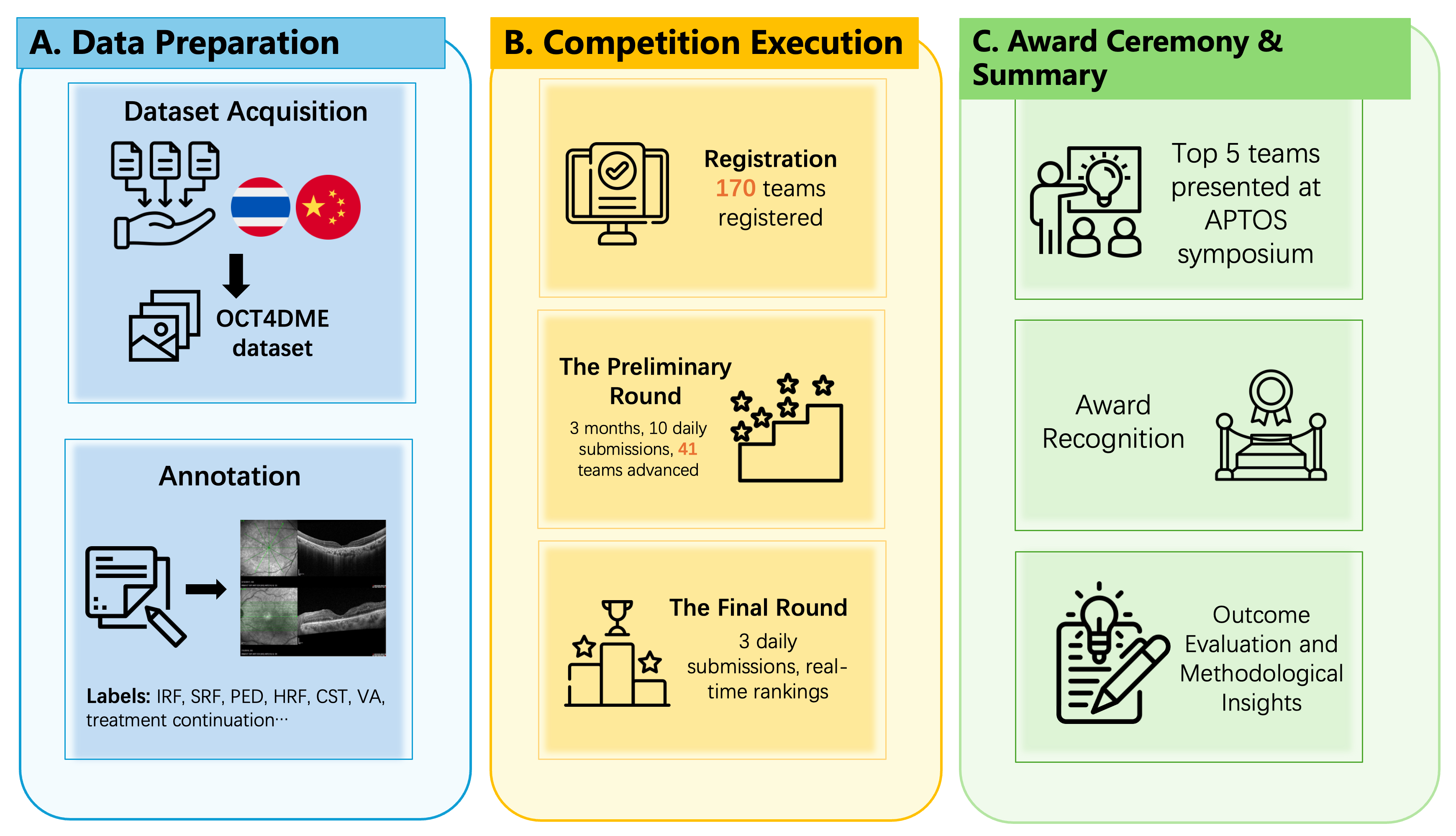}
\caption{The overall organization workflow of the 2021 Asia Pacific Tele-Ophthalmology Society (APTOS) Big Data Competition.} \label{overall}
\end{figure}

\subsection{Competition Overview}
The APTOS big data competition was officially launched from 28 Aug 2021 to 24 Jan 2022, as part of the 6th Asia-Pacific Tele-Ophthalmology Society (APTOS 2021) Symposium. The competition focused on predicting treatment outcomes for patients with DME undergoing anti-VEGF therapy based on OCT images. More details are available at \href{https://2021.asiateleophth.org/big-data-competition/}{https://2021.asiateleophth.org/big-data-competition/}. Figure \ref{overall} shows the overall competition organization workflow. 

Participants with the best scores on the private leaderboard were eligible for cash prizes: US\$8,000 for first place, US\$5,000 for second place, and US\$2,000 for third place. To qualify for the monetary awards, winning teams were required to present their solutions during the ``Share \& Tell'' session of the APTOS Monthly Webinar on February 12, 2022. Members affiliated with the organizing institutions were allowed to participate in the competition; however, they were not eligible to receive any awards.

The competition comprised three stages: the preliminary round, the final round, and the final presentation. In the preliminary round, participants accessed the training dataset and preliminary test dataset from the \href{https://tianchi.aliyun.com/dataset/120006}{APTOS website}. They were required to debug their algorithms locally and submit their results. This round spanned three months, during which each team had the opportunity to make up to 10 submissions per day. A total of 170 teams successfully submitted valid results to the challenge platform. The detailed timeline is as shown in Table \ref{tab:timeline}.

\begin{table}[htbp]
\centering
\caption{Timeline of the 2021 APTOS Big Data Competition.}
\begin{tabular}{|l|p{10cm}|}
\hline
\textbf{Date} & \textbf{Event} \\
\hline
August 28, 2021 & Launch of Competition, registration opens \\
\hline
September 1, 2021 & Submission portal opens; training and validation data released \\
\hline
November 19, 2021 & Entry deadline, team merger deadline, and preliminary round submission deadline \\
\hline
December 31, 2021 & Final submission deadline (Top 100 teams from the preliminary round eligible) \\
\hline
January 21, 2022 & Code defense: top 5 finalists undergo code review and panel challenge \\
\hline
February 12, 2022 & Prize presentation and winner showcase at the APTOS Monthly Webinar \\
\hline
\end{tabular}
\label{tab:timeline}
\end{table}

All the 41 teams from the preliminary round advanced to the final round. Teams that successfully reached the final round received links and decryption passwords to download the dataset. In the final round, each team was allowed up to 3 submissions per day, with real-time rankings. The top 5 teams participated in the final presentation in January 2022. During this concluding event, teams presented insights into their AI model development process, training techniques, and innovative ideas, ultimately determining the final rankings and awards for the competition.

\subsection{Competition Tasks}
The competition consisted of two rounds, each featuring distinct tasks focused on DME analysis using OCT. For detailed submission instructions, please refer to 
\href{https://www.alibabacloud.com/en/developer/competition/2nd-aptos-competition?_p_lc=1}{the official submission page}.

\subsubsection{The Preliminary Round:}

Participants were provided with OCT images both before and after treatment, along with the visual acuity (VA) before treatment. For each patient ID in the dataset, they were tasked with:
\begin{enumerate}
    \item Predicting the presence of Intraretinal Fluid (IRF), Subretinal Fluid (SRF), Pigment Epithelial Detachment (PED), and Hyperreflective Foci (HRF) in eyes.
    \item Predicting (measuring) the Central Subfield Thickness (CST) values before and after treatment.
    \item Predicting the VA after treatment.
    \item Predicting the continuation or discontinuation of anti-VEGF injection.
\end{enumerate}

\subsubsection{The Second Round:}

The tasks in the second round mirrored those in the first round. However, for the presence of IRF, SRF, PED, and HRF, participants were required to predict the probability (ranging from 0 to 1) rather than providing a simple yes/no response.

\subsection{Evaluation Metrics}

\subsubsection{The Preliminary Round:}

In this phase, participants utilize the first stage dataset and submit predictions for PreCST, CST, VA, Continue Injection (CI), IRF, SRF, and HRF. The scoring criteria are as follows:

\begin{enumerate}
    \item The area under the curve (AUC) is used for scoring predictions for Continue Injection, IRF, SRF, and HRF.
    \item For PreCST and CST, achieving predictions within the ±7.5\% confidence interval earns a score of 1.
    \item In the case of VA predictions:
    \\If the actual VA value is less than 1, a score of 1 is awarded for predictions within the range [VA-0.05, VA+0.05].
    \\If the actual VA value is greater than 1, a score of 1 is given for predictions within the range [VA×0.925, VA×1.075].
\end{enumerate}

Although predictions for preIRF, preSRF, preHED, preHRF, and HED are not directly evaluated, participants are still required to predict them. This is because predictions for continuing injection may heavily rely on improvements in these symptoms after treatment.

\subsubsection{The Final Round:}

The Final Round score incorporates participants' results on the second stage dataset. In other words, the score represents the average of the 14 indices submitted (7 indices from the first stage dataset and 7 indices from the second stage dataset). The leaderboard is sorted based on the Final Round score and displays sub-scores for each index to assist participants in optimizing their models.

\begin{figure}[t]
\includegraphics[width=\textwidth]{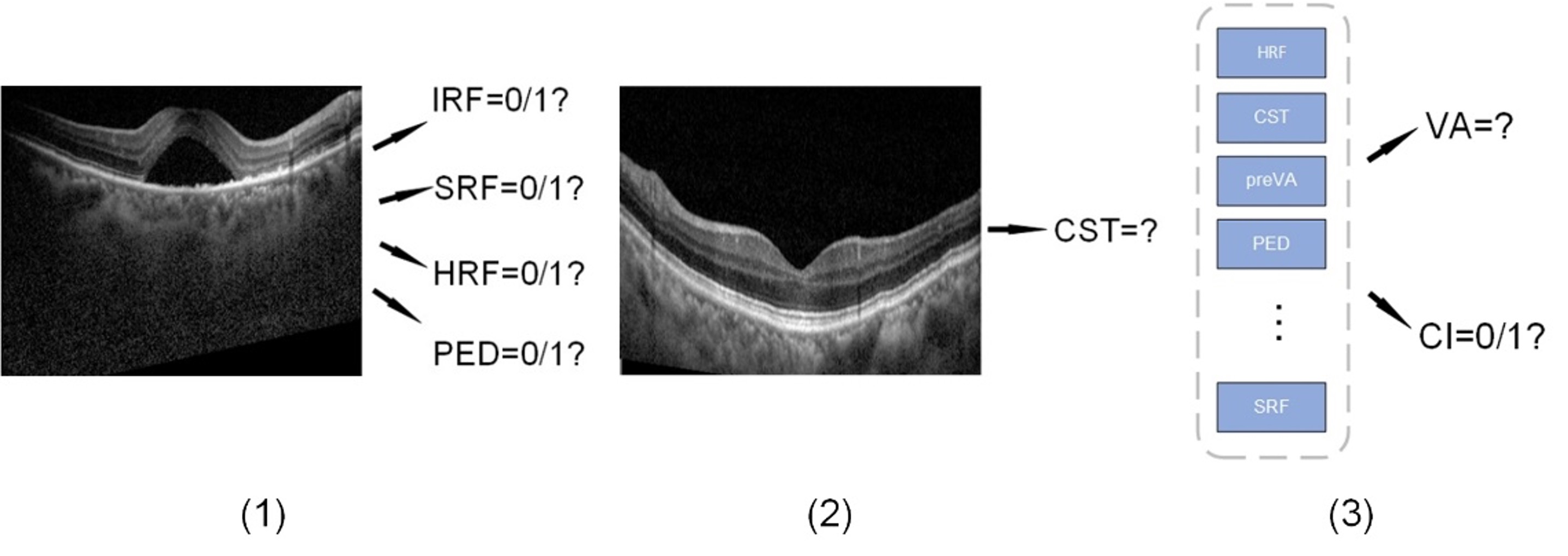}
\caption{Three sub-tasks of the competition, which are diagnosing the symptoms of the eye, predicting the Central Subfield Thickness, and judging whether to continue treatment.} \label{fig2}
\end{figure}

\section{Summary of Participating Solutions}
In this section, we compile the approaches employed by the top three winning teams in the competition. For ease of expression, we divided the 7 indices into three subtasks, as illustrated in Figure \ref{fig2}.

\subsection{Binary Classification Tasks for IRF, SRF, PED, and HRF}

\begin{figure}[t]
\includegraphics[width=\textwidth]{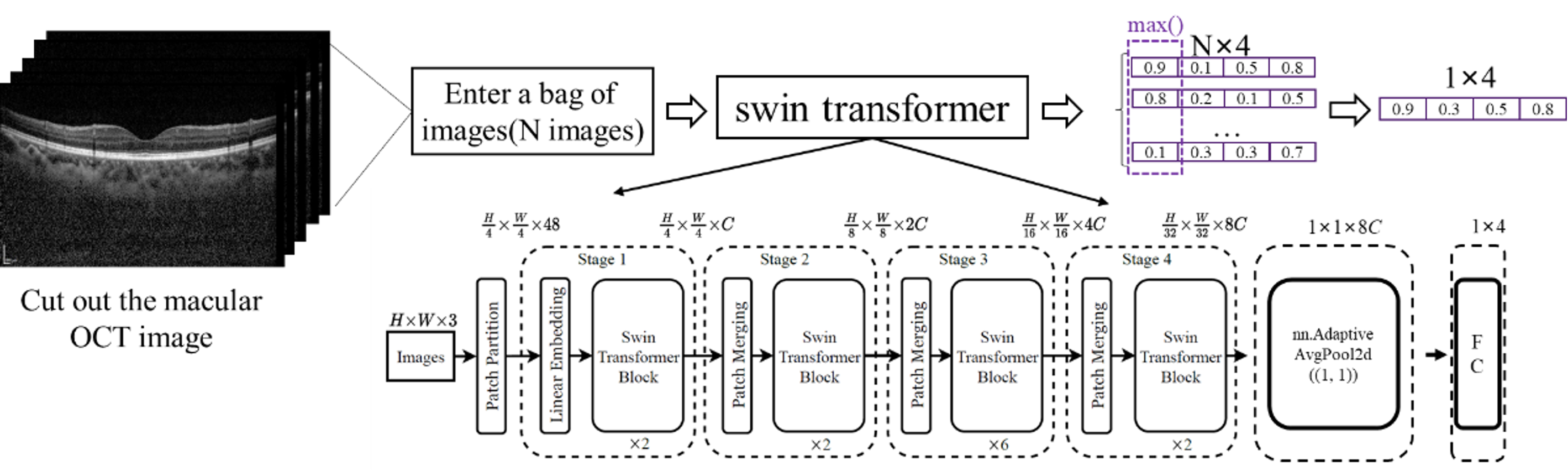}
\caption{Solution for Task 1 by the champion team BlueSky. This illustrates the architecture of the Multi-Instance Learning network proposed by the team. Given a bag containing n images, the network outputs an N x 4-dimensional vector, where 4 represents the categories: IRF, SRF, PED, and HRF.
} \label{fig3}
\end{figure}

\subsubsection{Team BlueSky:}
The champion team BlueSky encountered the challenge of transitioning from case-level to image-level labels within the dataset due to feature differences between the two-stage datasets. To address this hurdle, they employed a combination of Swin-Transformer \cite{RN18} and Multi-Instance Learning (MIL) \cite{RN19}. The team treated the classification task as an MIL problem, defining a "bag" as either encompassing all images within a case (for the first stage dataset) or each individual image (for the second stage dataset). For input, they extracted macular OCT image parts from the raw images. The Swin-Transformer was chosen as the backbone network, with a focus on macular OCT images and the utilization of max pooling to yield the final results. Additionally, they observed an imbalance in category distribution, particularly for IRF and HRF, between the first and the second stage datasets. Taking this into account, the team strategically balanced the category distribution in the merged dataset, contributing to an overall enhanced performance. The network architecture for this section is illustrated in Figure \ref{fig3}.

\subsubsection{Team LightRain:}
\begin{figure}[t]
\includegraphics[width=\textwidth]{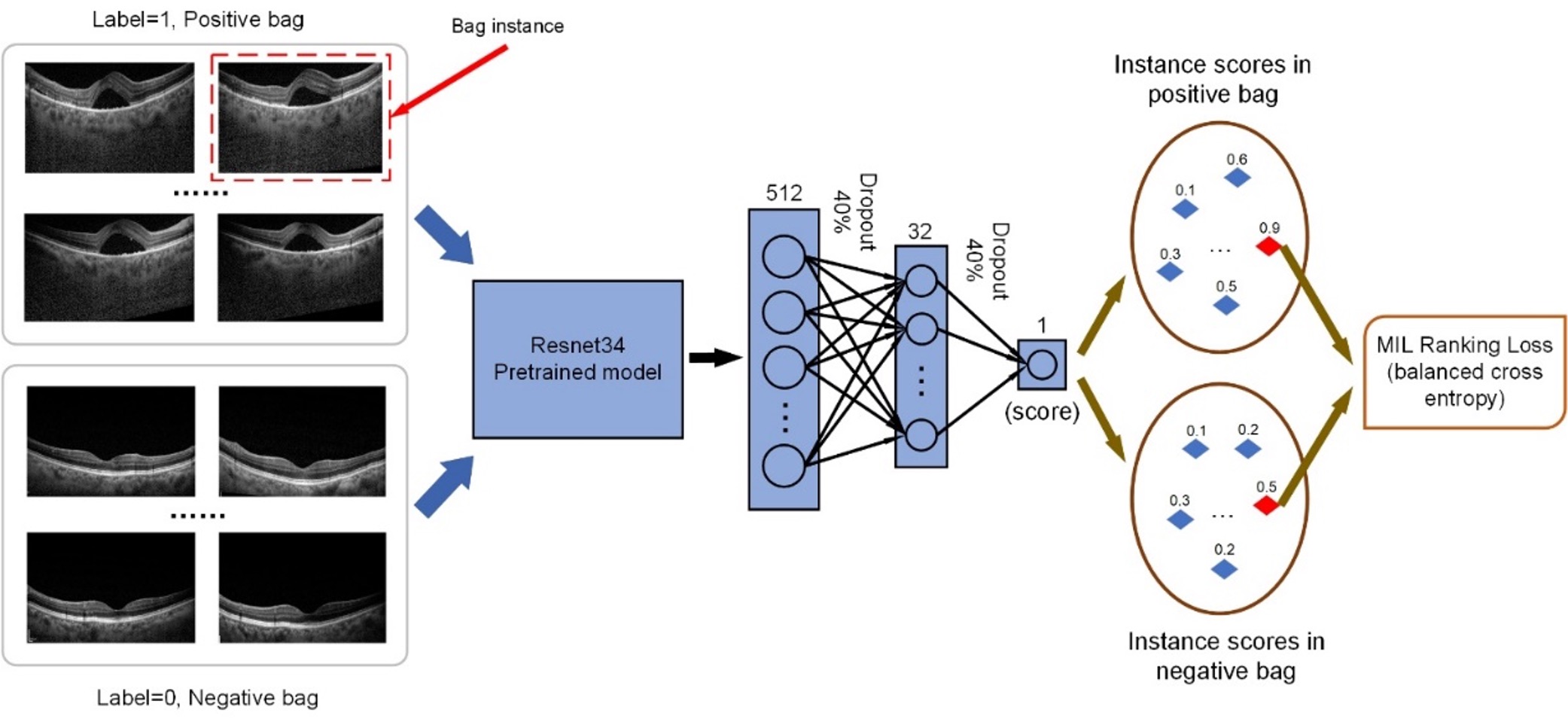}
\caption{Solution for Task 1 by the runner-up team LightRain. They used a weakly supervised learning framework for classification.} \label{fig4}
\end{figure}
 
The runner-up team LightRain addressed the imbalance between positive and negative samples using a weak supervision framework for MIL, as shown in Figure \ref{fig4}, similar to the approach employed by team BlueSky. They utilized ResNet34 \cite{RN20} as the base classifier and incorporated dropout to reduce overfitting. They introduced data augmentation to increase the diversity of training data, including random flipping, random cropping, and JPEG compression \cite{RN21}. The MIL Ranking Loss function \cite{RN22} and balanced cross-entropy loss were employed for optimization, so that only the samples with the highest score in the bag were used for optimization:
\begin{equation}
    L(B) = -\alpha y \log(\max_{i \in B} f(V_i)) - (1 - \alpha)(1 - y) \log(1 - \max_{i \in B} f(V_i)), \, \alpha \in (0, 1)
\end{equation}

The team also implemented Test Time Augmentation (TTA) \cite{RN23} for improved detection. 

\subsubsection{Team DarkStyle:}

\begin{figure}[t]
\includegraphics[width=\textwidth]{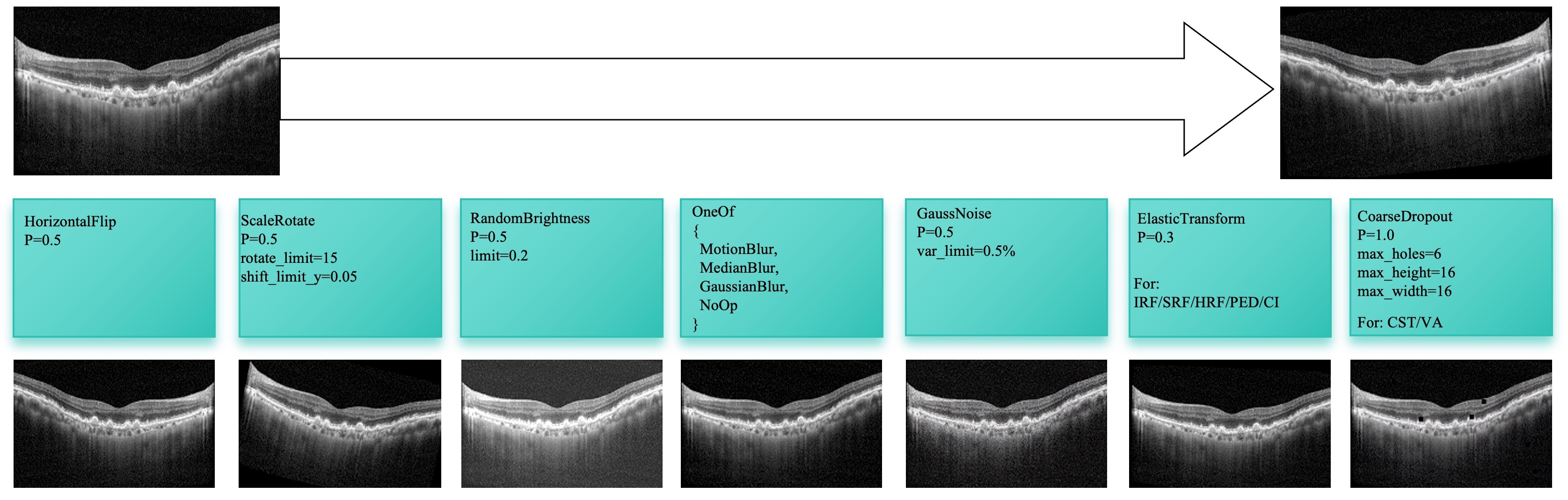}
\caption{The flow of data augmentation, proposed by the third-place team DarkStyle.} \label{fig5}
\end{figure}

Before conducting the experiments, the third-place team DarkStyle used an image augmentation flow to improve the generalization of the models, as shown in Figure \ref{fig5}. For the prediction of IRF/SRF/HRF/PED, they utilized EfficientNet-B0 \cite{RN24} as the backbone. A global average pool layer and multilayer perceptron layers were added, and binary cross-entropy loss with label smoothing set to 0.1 was employed to combat overfitting. For each task, they trained a separate model, resulting in a total of four classification models. Particularly, in the prediction of PED, they conducted a gamma transform for image preprocessing. The optimization used the Adam optimizer with a learning rate of 0.001. The team experimented with various backbones, including ResNet50 \cite{RN20}, MobileNetV2 \cite{RN25}, and EfficientNet-B0 \cite{RN24} in the preliminary round, with EfficientNet-B0 demonstrating the best balance between accuracy and efficiency. Their approach included rigorous validation techniques including the 5-fold cross-validation to ensure robust model performance.

\subsection{Regression Tasks for Central Subfield Thickness (CST)}

\subsubsection{Team BlueSky:}

In tackling the regression task of CST values before and after treatment, the team BlueSky faced the complexities arising from diverse scanning methods (radial and horizontal) and data irregularities. They devised a robust approach for identifying the central image by computing the sum of rows of the G channel in fundus images. For CST regression, they opted for ResNet101 \cite{RN20} as the backbone, and introduced a tailored loss function that integrated mean absolute percentage error with smooth L1 loss:

\begin{subequations}
\begin{equation}
    M = \left| \frac{\hat{y_i} - y_i}{y_i} \right|
\end{equation}

\begin{equation}
    \text{loss} = 
    \begin{cases} 
    0.5 \times \frac{M^2}{\beta}, & M < \beta, \quad \beta = 0.075 \\
    M - 0.5 \times \beta, & M \geq \beta, \quad \beta = 0.075 
    \end{cases}
\end{equation}

\end{subequations}

\begin{figure}[t]
\centering
\includegraphics[width=0.85\textwidth]{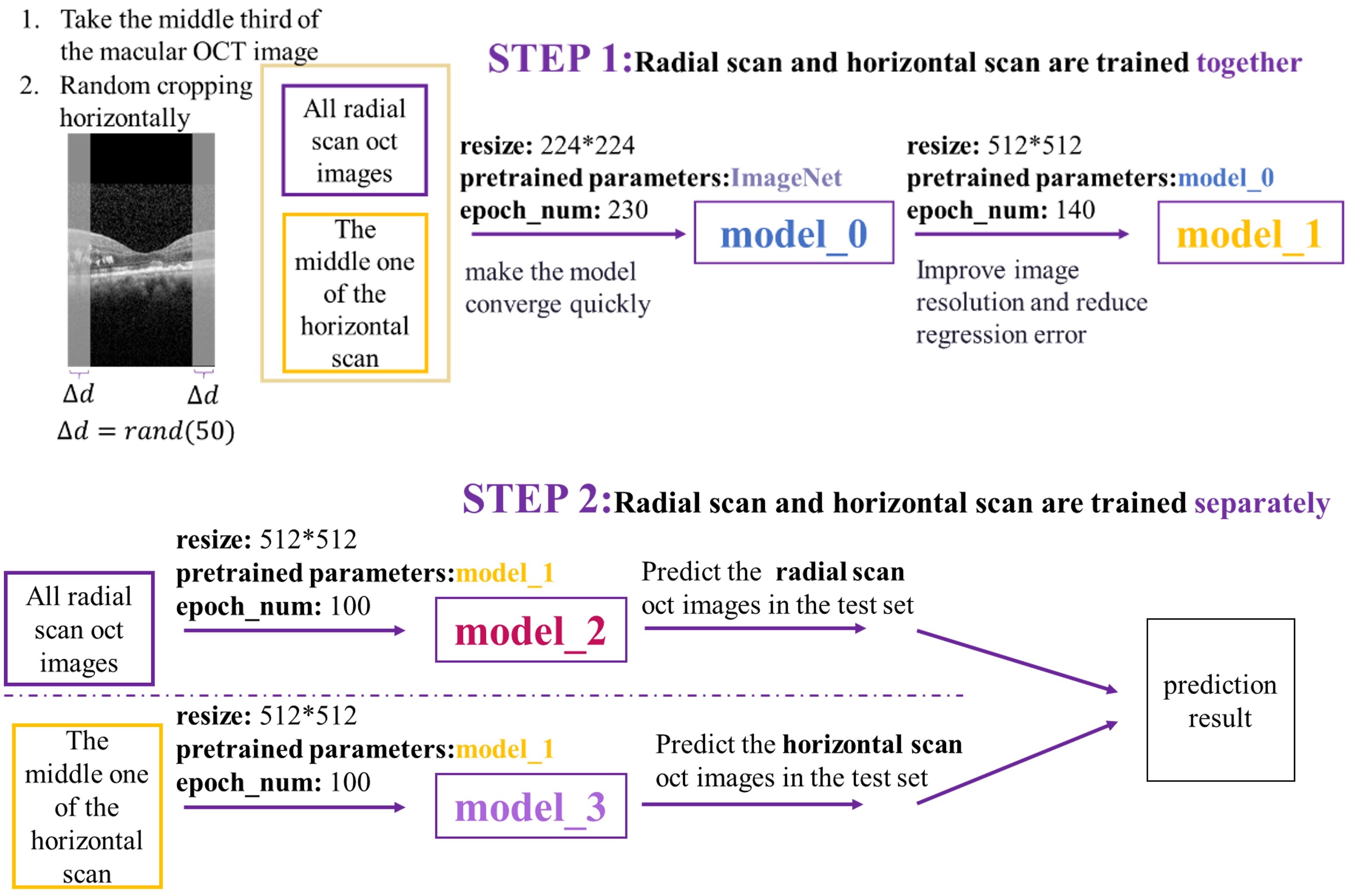}
\caption{Solution for Task 2 by the champion team BlueSky. They used a two-step training progress for Central Subfield Thickness (CST) regression.} \label{fig6}
\end{figure}

During data input, they focused on the central third of the macular OCT image and performed random cropping within a range of 50 pixels horizontally. To improve model generalization, the training process encompassed distinct stages for radial and horizontal scans. The ultimate model was refined through iterative training steps, as illuminated in Figure \ref{fig6}, culminating in a robust and computationally efficient solution. In the initial phase, both radial and horizontal scans were initially trained together to ensure an ample sample size. Team BlueSky resized the image to 224, employed Imagenet \cite{RN26} pre-training parameters, and trained for 230 epochs to achieve quick convergence, resulting in model 0. Subsequently, they resized the image to 512, used the parameters from model 0, and trained for 140 epochs to enhance image resolution and reduce regression errors, yielding model 1. In the second phase, they conducted separate training for radial and horizontal scans. Using an image size of 512 and the parameters from model 1, they trained the two scans independently. The predictions from the two scans were then combined to generate the final result.

\subsubsection{Team LightRain:}

For the regression task on CST, the runner-up team LightRain utilized Densenet169 as the backbone \cite{RN27}. They encountered challenges related to the quality of labeled data. To address this, they implemented a refinement strategy: during a preliminary test, samples with substantial discrepancies (exceeding 15\%) between the predicted and actual values were identified and removed from the training set. The CST values were normalized, and the Mean Absolute Error (MAE) loss served as the loss function. Random inversion was applied as a data augmentation technique to mitigate CST’s sensitivity to target size.

\subsubsection{Team DarkStyle:}

For CST regression, the third-place team DarkStyle applied a key strategy to improve the performance by excluding horizontal scan images far from the center point. This involved analyzing the left part of the original OCT images and applying distance and selection criteria to filter out relevant images. The optimal parameters for training and prediction were determined separately for the images from the first and the second stage datasets. EfficientNet-B0 was utilized as the backbone. 
\subsection{Prediction of Patients' Visual Acuity and Treatment Continuation}
\begin{figure}[t]
\centering
\includegraphics[width=0.95\textwidth]{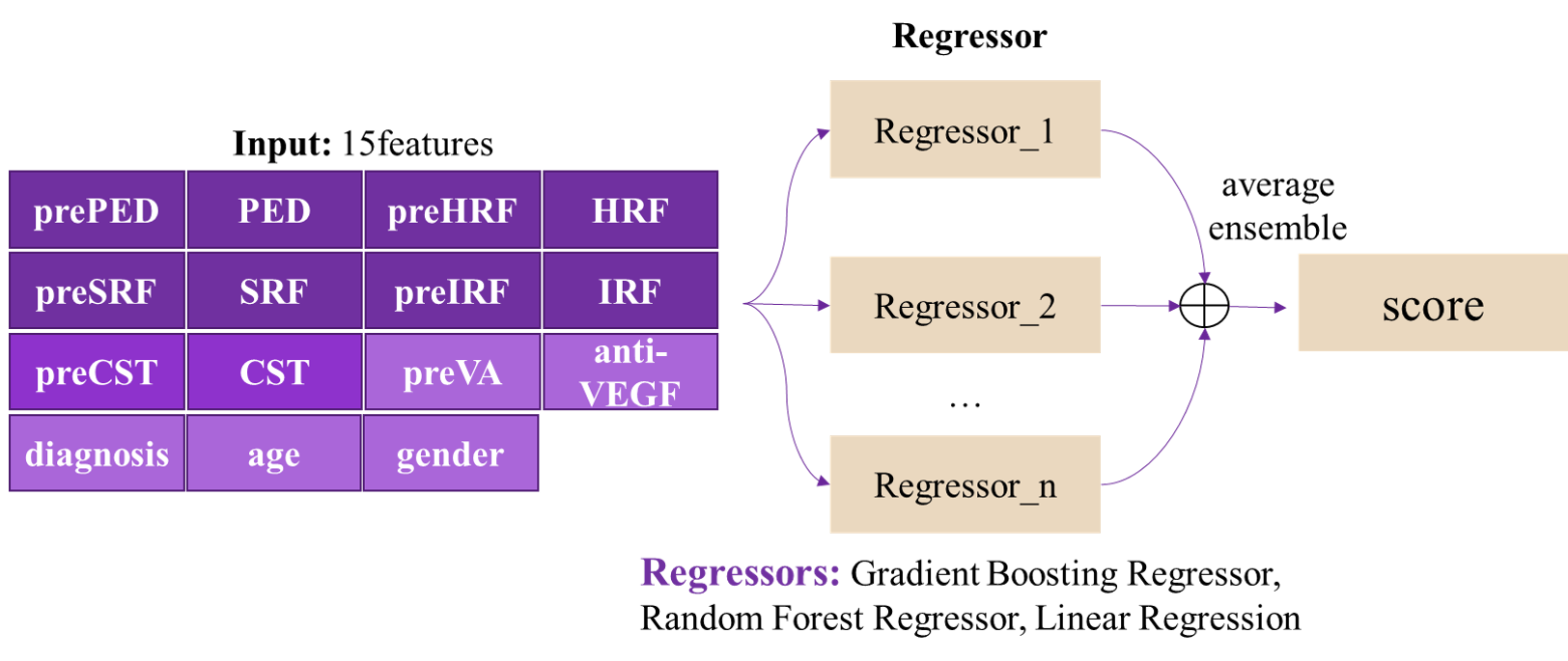}
\caption{The ensemble strategy for predicting visual acuity and treatment continuation proposed by the champion team BlueSky.} \label{fig7}
\end{figure}
\subsubsection{Team BlueSky:}
In predicting patients' VA and treatment continuation, BlueSky integrated the outcomes from the previous tasks. Recognizing the potential impact of errors in both classification and regression tasks, they implemented a robust approach using a voting regressor for enhanced stability, as shown in Figure \ref{fig7}. The decision-making process considered 15 features, including classification results, regression results, and sponsor-provided features. The ensemble strategy involved multiple regressors, such as Gradient Boosting Regressor \cite{RN28}, Random Forest Regressor \cite{RN29}, and Linear Regression, contributing to the overall stability and reliability of the model.

\subsubsection{Team LightRain:}
The team LightRain also combined results from the first two tasks with patient information to form the input feature vectors. For preprocessing, they conducted one hot encoding for the disease types and injection drug types. A three-layer neural network was utilized to predict VA changes, and an integrated learning framework Gradient Boosting was applied to predict whether patients need continuous injection therapy.

\begin{figure}[t]
\includegraphics[width=\textwidth]{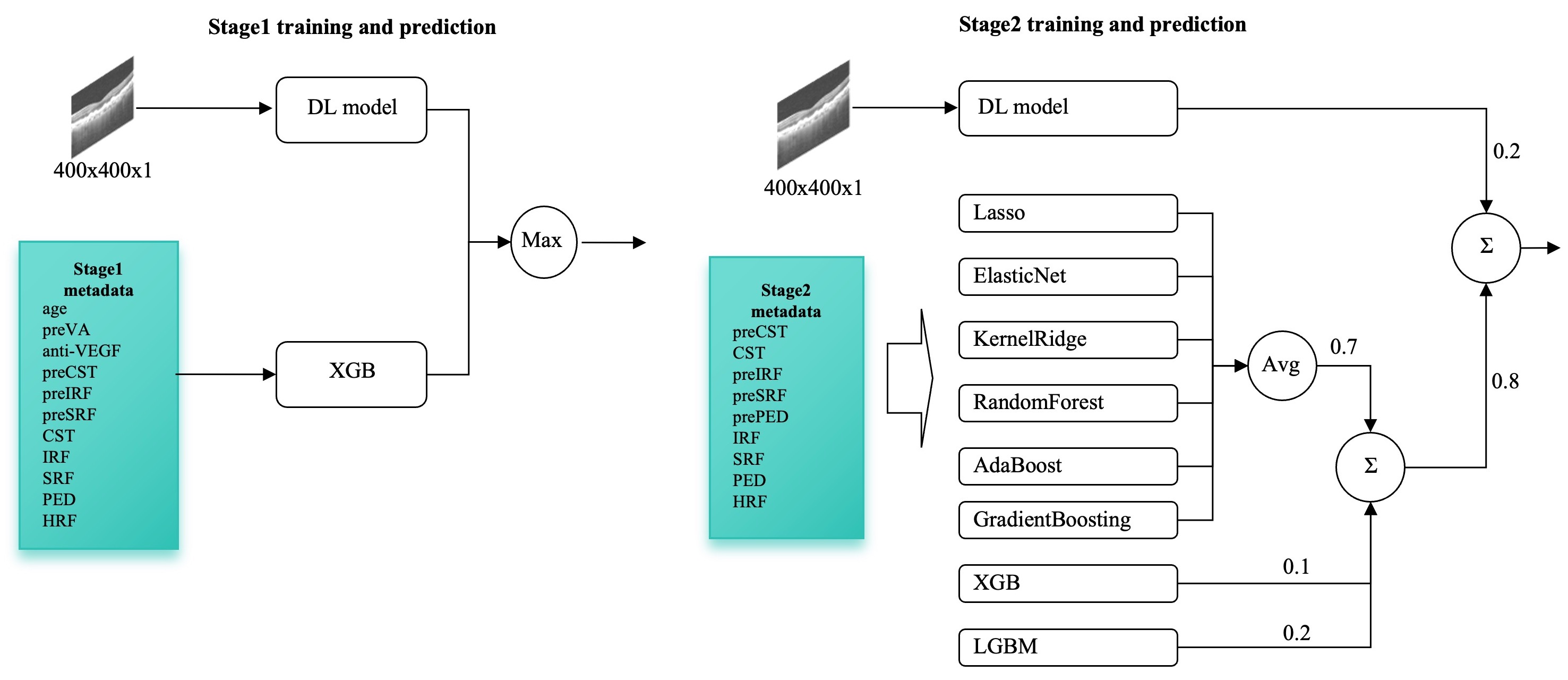}
\caption{Solution for predicting continued injection by the third-place team DarkStyle.  They used a two-step training progress for Central Subfield Thickness (CST) regression.} \label{fig8}
\end{figure}

\subsubsection{Team DarkStyle:} 
The third-place team approached VA prediction as a classification task for the first-stage data and a regression task for the second-stage data. Both stages jointly used image features and patient metadata. To predict the need for continued injections, this team performed an in-depth analysis of metadata, dropping low correlation columns such as gender. They leveraged the diagnosis value equal to 6 as an indicator of a fellow eye, transforming CST values accordingly. The overall architecture is illustrated in Figure \ref{fig8}. The team explored simple patterns based on observations in training metadata and introduced a deep learning model for image prediction and an XGBoost Regression model for metadata prediction in the first stage. For the second stage, eight different models were used for metadata, and the results were weighted and combined to enhance accuracy.

\begin{figure}[t]
\includegraphics[width=\textwidth]{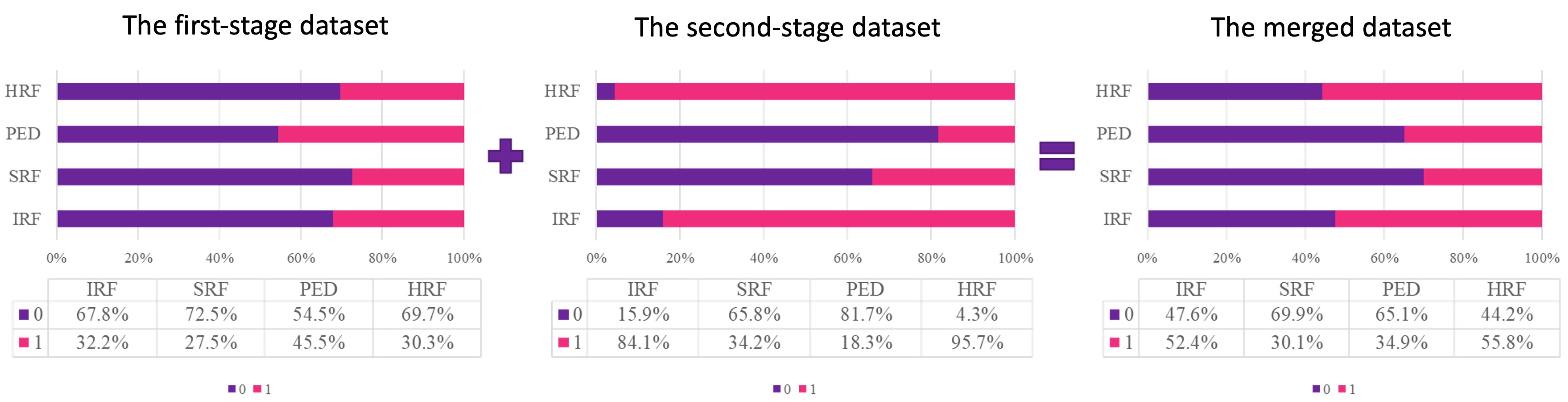}
\caption{The distribution of data for IRF (Intraretinal Fluid), SRF (Subretinal Fluid), PED (Pigment Epithelium Detachment), and HRF (Hyperreflective Foci) in datasets at different stages.} \label{fig9}
\end{figure}

\section{Results}
\subsection{Data Distribution}
The disparity in category distribution between the preliminary and final datasets, particularly in the cases of IRF and HRF, raises interesting considerations. Figure \ref{fig9} reveals a substantial shift in the proportions, with IRF increasing from 32.2\% in the first-stage dataset to 84.1\% in the second stage, and finally stabilizing at 52.4\% after the combination. Likewise, HRF starts at 30.3\% in the first stage, surges to 95.7\% in the second stage, and then settles at 55.8\% post-combination. This observed shift highlights the impact of dataset merging on achieving a more balanced sample distribution. It prompts exploration into the implications of category distribution dynamics on the performance of models across different stages of the competition.

\subsection{Overall Results}

Table \ref{tab3} displays the top ten teams based on their overall scores. We present their performances in both the preliminary round (evaluation on the first stage test set) and the final round (evaluation on the merged test set). A noteworthy observation is the substantial increase in scores for all teams from the preliminary round to the final round. Teams including BlueSky, GuardianHairline, c\_c, HowToImproveScore, and UnityTeam maintain or improve their rankings on the test dataset, indicating their robustness compared to other methods. Conversely, teams with lower rankings in the final round generally exhibit poorer generalization capabilities. Specifically, LightRain, DarkStyle, ChubbyFirst, MedEngTogether, and MedImgProc experienced a decline in rankings on the final test set.

\begin{table}[t]
\caption{Grading results for the 2021 APTOS Big Data Competition. Teams are ranked based on the overall score in the final round. The numbers in parentheses after the scores indicate the team's ranking in the preliminary round, with bolded numbers indicating a rise or no change in the ranking between the preliminary and final rounds, and non-bolded numbers indicating a fall in the rankings.}
\centering
\resizebox{\textwidth}{!}{%
\begin{tabular}{lllll}
\hline
\multicolumn{1}{c}{\textbf{Rank}} & \textbf{Team}     & \textbf{Organization}                                 & \textbf{Preliminary} & \textbf{Final} \\ \hline
1                                 & BlueSky           &                                                       & 0.53 (12)            & 0.7440         \\
2                                 & LightRain         & Xi'an Electronic Science and Technology University    & 0.56 (1)             & 0.7399         \\
3                                 & DarkStyle         & Visual Microimaging                                   & 0.56 (1)             & 0.7354         \\
4                                 & ChubbyFirst       &                                                       & 0.56 (1)             & 0.7332         \\
5                                 & GuardianHairline  & Hangzhou Electronic Science and Technology University & 0.46 (31)            & 0.7323         \\
6                                 & c\_c              & Chengdu University of Information Technology          & 0.54 (10)            & 0.7296         \\
7                                 & MedEngTogether    & Tsinghua University                                   & 0.56 (1)             & 0.7261         \\
8                                 & HowToImproveScore & Beijing University of Technology                      & 0.55 (8)             & 0.7228         \\
9                                 & MedImgProc        & South China University of Technology                  & 0.55 (8)             & 0.7166         \\
10                                & UnityTeam         & Visual Microimaging                                   & 0.44 (50)            & 0.7084         \\ \hline
\end{tabular}
}
\label{tab3}
\end{table}

\subsection{Individual Results}
The detailed results of the top 3 teams for each subtask are demonstrated in Table \ref{tab4}. The champion team, BlueSky, showcased a remarkable overall performance with a score of 0.7440. Particularly noteworthy was their strength in CST during the first round, where they achieved a high score of 0.7026, demonstrating the effectiveness of their regression approach.

\begin{table}[t]
\caption{Detailed results of the top 3 teams for each subtask. The scores with the suffix "\_stage1" represent the performance of the teams on the first-stage dataset, while the scores with the suffix "\_stage2" indicate the teams' performance on the second-stage dataset.}
\centering
\resizebox{\textwidth}{!}{
\begin{tabular}{lllllllllllllllll}
\hline
\textbf{Rank} &
  \textbf{Team} &
  \textbf{Score} &
  \textbf{preCST\_stage1} &
  \textbf{VA\_stage1} &
  \textbf{CI\_stage1} &
  \textbf{CST\_stage1} &
  \textbf{IRF\_stage1} &
  \textbf{SRF\_stage1} &
  \textbf{HRF\_stage1} &
  \textbf{preCST\_stage2} &
  \textbf{VA\_stage2} &
  \textbf{CI\_stage2} &
  \textbf{CST\_stage2} &
  \textbf{IRF\_stage2} &
  \textbf{SRF\_stage2} &
  \textbf{PED\_stage2} \\ \hline
1 & BlueSky   & 0.7440 & 0.5643 & 0.3216 & 0.7026 & 0.5906 & 0.9293 & 0.9218 & 0.8593 & 0.6871 & 0.5234 & 0.7828 & 0.6930 & 0.9338 & 0.9264 & 0.9796 \\
2 & LightRain & 0.7399 & 0.5468 & 0.3216 & 0.6852 & 0.5994 & 0.9221 & 0.9311 & 0.8772 & 0.6462 & 0.5234 & 0.7800 & 0.6842 & 0.9313 & 0.9300 & 0.9796 \\
3 & DarkStyle & 0.7354 & 0.5965 & 0.3421 & 0.6983 & 0.5994 & 0.9058 & 0.8854 & 0.8485 & 0.6520 & 0.5234 & 0.8006 & 0.643  & 0.9074 & 0.9174 & 0.9749 \\ \hline
\end{tabular}
}
\label{tab4}
\end{table}
The runner-up team, LightRain, achieved a commendable score of 0.7399. They exhibited strength in predicting whether to Continue Injection or not in the second stage with a score of 0.7800, highlighting the high clinical applicability of their method. Furthermore, they maintained high scores in the first stage SRF (0.9311) and the first stage IRF (0.9221), indicating their excellent performance in classification tasks.

The third-place team, DarkStyle, secured a score of 0.7354 and demonstrated noteworthy strengths in the second stage prediction. They predicted the second stage CST with a score of 0.8006. They also displayed strong results in the second stage SRF (0.9174) and in the second stage IRF (0.9074), contributing to their solid overall performance and ranking.

Overall, each team displayed unique strengths in various aspects of the competition. BlueSky excelled in CST and SRF, LightRain showcased competitiveness in CI, and DarkStyle demonstrated strengths in CST and SRF. These three winning teams maintained a commendable overall performance throughout the competition, highlighting their consistency and expertise.

The results showcase exceptional performance in various tasks: The decision consistency for Continued Injection is notable, with Al aligning with ophthalmologists in 80.06\% of the cases. In PED detection, an impressive AUC of 97.96\% was achieved. SRF detection and IRF detection exhibited high AUC rates of 93.68\% and 93.38\%, respectively. For CST prediction before treatment and after treatment, the model achieved AUCs of 68.71\% and 69.30\%, respectively. VA prediction after treatment maintained a commendable AUC of 55.56\%.

In summary, the models developed by the participants exhibited outstanding performance across a range of crucial parameters, underscoring their versatility and effectiveness in tackling intricate OCT tasks. This underscores the potential of utilizing machine learning to address the challenges associated with DME, particularly in predicting patient responses before treatment, enabling personalized treatment plans, and improving patient management.

\section{Discussions}
In this section, we summarize the characteristics, experiences, and insights gained from the methods employed during the competition. Additionally, we analyze the value of artificial intelligence and machine learning approaches in DME intervention, along with feasible future works.

\subsection{Data preprocessing and analysis}

A solid foundation lies in data analysis. The top two teams conducted meticulous analyses of the dataset distribution before training and took steps to equalize sample distribution. Balanced datasets ensure that the model does not develop biases toward more frequent classes, improving both the generalizability and reliability of predictions. This is particularly important in medical applications, where model errors could directly impact patient care and treatment plans. The runner-up team LightRain implemented a two-stage training approach to pre-filter noise samples, ensuring enhanced training quality.

To maximize data utilization, all winning teams designed their data augmentation methods, tailoring them for different tasks. The third-place team DarkStyle introduced seven distinct augmentation techniques, creating an augmentation flow for diverse and enriched training data, as in Figure \ref{fig5}.

Notably, the avoidance of image compression in regression tasks and the preference for larger image sizes, especially in medical datasets, emerge as valuable insights gleaned from the competition.

\subsection{Model Innovations and Generalization}
The competition results demonstrates that Models need not be overly complex; instead, employing general deep learning models proves effective in addressing the competition tasks. Successful teams opted for commonly used deep learning models to tackle the challenges. For instance, the team LightRain achieved outstanding results by utilizing a lighter-weight model (ResNet35) for Binary Classification Tasks related to IRF, SRF, PED, and HRF. Despite the simplicity, their approach ranked first in the preliminary round and second in the final.

In the realm of medical image analysis, the design of task-specific models often outperforms the pursuit of complex and novel model architectures. Participating teams conducted a thorough analysis of the requirements for various subtasks in the competition. For tasks like IRF/SRF/PED/HRF classification, a chosen approach involved using classification methods, while tasks involving CST numerical predictions favored regression methods. Key comprehensive decision-making tasks, such as predicting VA after treatment and determining the need for continued injection, found Ensemble methods to be more robust. Ensemble methods can amalgamate judgments from multiple models to make a final decision. These models combine the insights from multiple models to reach a final decision, enhancing overall accuracy and reliability. Additionally, employing multi-stage models proves advantageous, allowing for detailed parameter adjustments and optimization. Both the team BlueSky and team LightRain adopted similar multi-stage approaches.

\subsection{Clinical Findings and Implications}

\begin{figure}[t]
\centering
\includegraphics[width=0.8\textwidth]{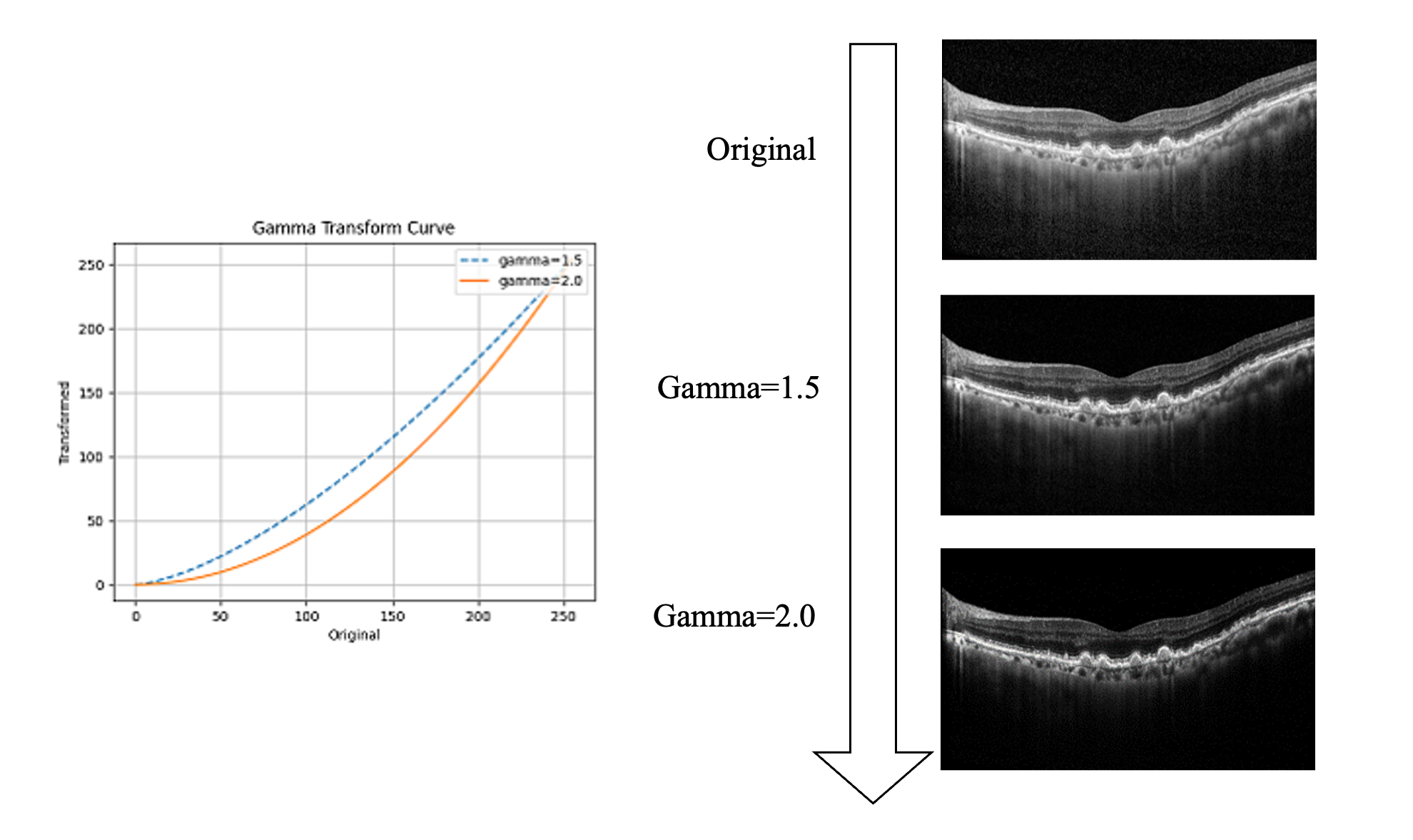}
\caption{Gama transform of the OCT images.} \label{fig10}
\end{figure}

In the context of predicting PED, the team DarkStyle observed a notable enhancement in the AUC by approximately 0.3\% with the application of a gamma transform. The gamma transform, characterized by a transform coefficient greater than 1.0, accentuates dark pixels while dimming light pixels, as illustrated in Figure \ref{fig10}. This improvement can be attributed to that PED symptoms predominantly manifest in the bottom-lightest part of the retina. By applying the gamma transform, these subtle features become more distinguishable, facilitating more accurate identification of PED.

\begin{figure}[t]
\includegraphics[width=\textwidth]{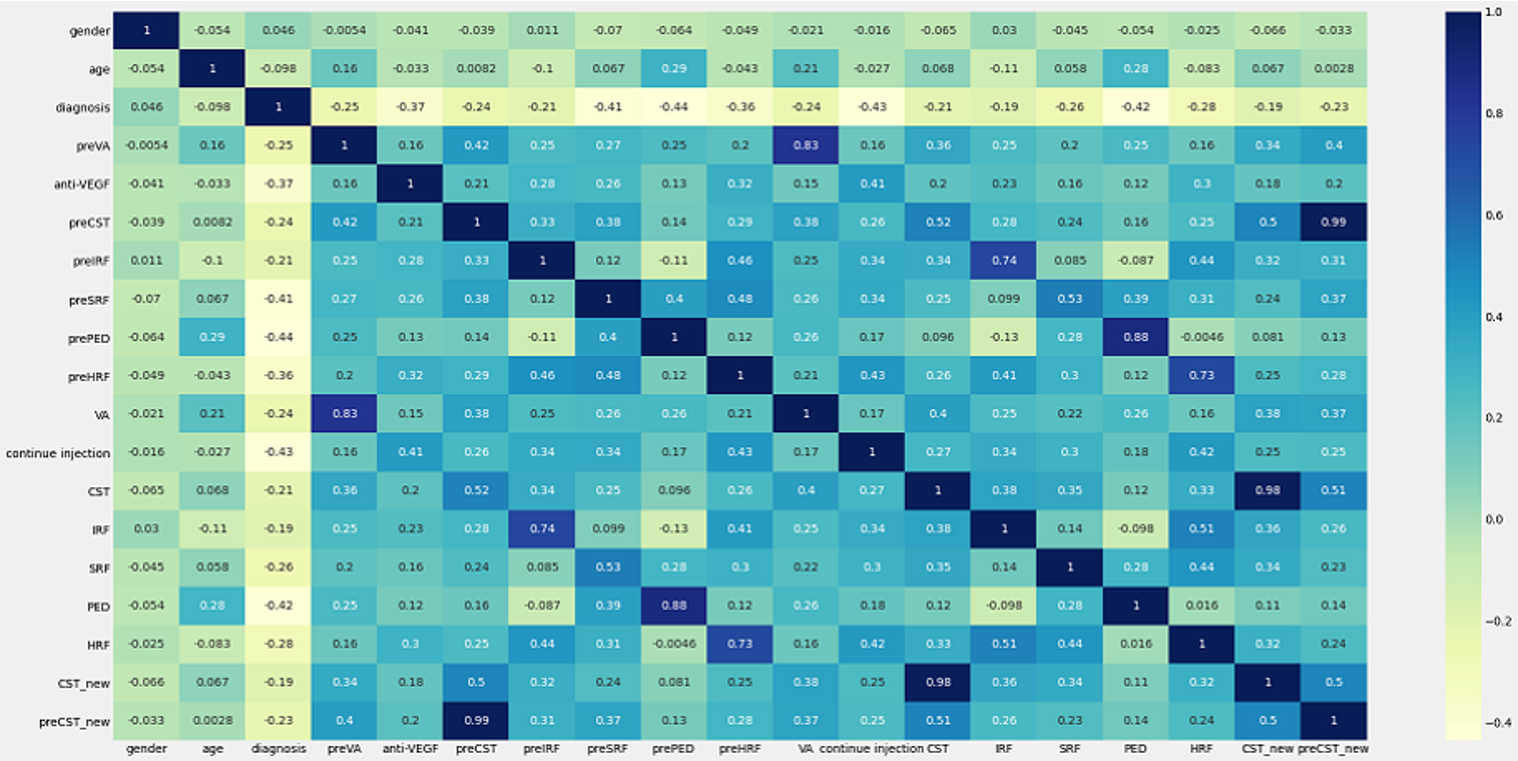}
\caption{The correlation matrix for all metadata columns in the dataset.} \label{fig11}
\end{figure}

To unravel inter-variable relationships, a correlation matrix was plotted for all metadata columns, and low correlation columns, such as gender, were subsequently dropped, as depicted in Figure \ref{fig11}. Notably, a diagnosis value of 6 corresponds to the fellow eye, signifying a healthy eye or one not severe enough to necessitate treatment. Analysis revealed that the mean value of the fellow eye's CST is approximately 250, suggesting deviations from this value may indicate anomalies. Thus, the original CST values were transformed by subtracting 250.

Further exploration involved seeking simple patterns based on training metadata observations, such as:
\begin{enumerate}
    \item If IRF, SRF, HRF, and PED all disappeared after treatment, does this imply that continuing injections is entirely unnecessary? Contrary to this assumption, 38 cases in the first-stage training data required continued injections, indicating that these features alone are not sufficient for decision-making.
    \item Similarly, if IRF, SRF, HRF, and PED all appeared after treatment, does it signify a 100\% unnecessary continuation of injection, suggesting the treatment had no effect? Again, the answer is negative, with 44 continued injection cases found in the first-stage training data.
\end{enumerate}
Exploration extended to more intricate assumptions, including changes in VA, yet no straightforward rules were discovered. This suggests that crucial information influencing the decision to continue injection is embedded in the OCT images, beyond what can be inferred from simple metadata patterns. Future work may benefit from combining metadata analysis with deeper image feature extraction to improve treatment decision-making.

\subsection{Advancing Towards Clinical Application}

A pivotal question emerges: Can these models be effectively employed in real clinical settings to autonomously predict whether patients with DME should undergo anti-VEGF treatment? While this question remains open, the development of an automated, objective diagnostic system holds promise in mitigating individual biases and conserving significant time for human experts. OCT, being non-invasive and cost-efficient, is extensively utilized by clinical experts for DME monitoring \cite{RN30}. Leveraging AI methods with OCT for automated anti-VEGF therapy decisions in patients with DME appears to be a viable solution for assisting diagnosis under limited conditions \cite{RN31,RN32}. In our competition, these models exhibit impressive accuracies, reaching up to an AUC of 80.06\%, although the results are specific to the Asian population, with data collected from Thailand, India, and China.

One challenge is that the lack of explainability may raise concerns among medical professionals, making them hesitant to use AI-based models. In future endeavors, we plan to explore the interpretability of machine learning methods more thoroughly. For instance, visualizing the regions of interest in OCT by Integrated Gradients \cite{RN33} would allow us to analyze the basis of AI models when making predictions using OCT. Furthermore, we aim to validate the effectiveness of AI models in assisting anti-VEGF injection diagnosis beyond the Asian population, enhancing the generalizability of our conclusions. Additionally, prior to the formal integration of these methods into clinical studies, it is imperative to identify and address potential limitations, including data quality, integration with screening workflows, and potential biases. Rigorous monitoring and evaluation of the performance and safety of these AI models post-regulatory approval are also essential.

\section{Conclusion}
We organized a competition focused on DME analysis using OCT data, with the key aim of predicting patient responses to treatment, specifically the decision of whether to continue anti-VEGF injection therapy. The competition addresses the unique challenges of DME analysis through OCT. This paper presents the dataset we released tailored for this task, outlines the competition's process and evaluation framework, and presents top-performing algorithms. Offering the first in-depth exploration of OCT multi-task DME analysis, our work lays the foundation for further research in this clinically relevant domain. Researchers are encouraged to leverage the openly accessible dataset to advance the field further.

\section*{Declaration of Competing Interest}

The authors declare that they have no known competing financial interests or personal relationships that could have appeared to influence the work reported in this paper. 


\section*{Data Availability}
The dataset is available to the public on the official website at \href{https://tianchi.aliyun.com/dataset/120006}{APTOS2021 Dataset}. Interested applicants are required to fill out the application form on the \href{https://asiateleophth.org/credential-application/}{APTOS website}, and sign the Data Use Agreement. The application will undergo review, and APTOS will respond within 7 working days.

\section*{Fundings}
The study was supported by the Global STEM Professorship Scheme (P0046113) and Henry G. Leong Endowed Professorship in Elderly Vision Health.

\section*{Acknowledgments} 
We thank the InnoHK HKSAR Government for providing valuable supports.

This research received support from the JC STEM Lab of Innovative Light Therapy for Eye Diseases funded by The Hong Kong Jockey Club Charities Trust. The sponsor or funding organization did not participate in the design or implementation of this study.
\pagebreak
\bibliographystyle{elsarticle-num} 
\bibliography{Manuscript.bib}

\begin{thebibliography}{10}
\expandafter\ifx\csname url\endcsname\relax
  \def\url#1{\texttt{#1}}\fi
\expandafter\ifx\csname urlprefix\endcsname\relax\def\urlprefix{URL }\fi
\expandafter\ifx\csname href\endcsname\relax
  \def\href#1#2{#2} \def\path#1{#1}\fi

\bibitem{RN1}
H.~Hashemi, F.~Rezvan, R.~Pakzad, A.~Ansaripour, S.~Heydarian, A.~Yekta, H.~Ostadimoghaddam, M.~Pakbin, M.~Khabazkhoob, Global and regional prevalence of diabetic retinopathy; a comprehensive systematic review and meta-analysis, Semin Ophthalmol 37~(3) (2022) 291--306.
\newblock \href {https://doi.org/10.1080/08820538.2021.1962920} {\path{doi:10.1080/08820538.2021.1962920}}.

\bibitem{RN2}
A.~Daruich, A.~Matet, A.~Moulin, L.~Kowalczuk, M.~Nicolas, A.~Sellam, P.~R. Rothschild, S.~Omri, E.~Gélizé, L.~Jonet, K.~Delaunay, Y.~De~Kozak, M.~Berdugo, M.~Zhao, P.~Crisanti, F.~Behar-Cohen, Mechanisms of macular edema: Beyond the surface, Prog Retin Eye Res 63 (2018) 20--68.
\newblock \href {https://doi.org/10.1016/j.preteyeres.2017.10.006} {\path{doi:10.1016/j.preteyeres.2017.10.006}}.

\bibitem{RN3}
S.~D. Solomon, E.~Chew, E.~J. Duh, L.~Sobrin, J.~K. Sun, B.~L. VanderBeek, C.~C. Wykoff, T.~W. Gardner, Diabetic retinopathy: A position statement by the american diabetes association, Diabetes Care 40~(3) (2017) 412--418.
\newblock \href {https://doi.org/10.2337/dc16-2641} {\path{doi:10.2337/dc16-2641}}.

\bibitem{RN4}
J.~Zhang, J.~Zhang, C.~Zhang, J.~Zhang, L.~Gu, D.~Luo, Q.~Qiu, Diabetic macular edema: Current understanding, molecular mechanisms and therapeutic implications, Cells 11~(21) (2022).
\newblock \href {https://doi.org/10.3390/cells11213362} {\path{doi:10.3390/cells11213362}}.

\bibitem{RN5}
N.~M. Bressler, W.~T. Beaulieu, A.~R. Glassman, K.~J. Blinder, S.~B. Bressler, L.~M. Jampol, M.~Melia, r.~Wells, J.~A., Persistent macular thickening following intravitreous aflibercept, bevacizumab, or ranibizumab for central-involved diabetic macular edema with vision impairment: A secondary analysis of a randomized clinical trial, JAMA Ophthalmol 136~(3) (2018) 257--269.
\newblock \href {https://doi.org/10.1001/jamaophthalmol.2017.6565} {\path{doi:10.1001/jamaophthalmol.2017.6565}}.

\bibitem{RN6}
J.~Olson, P.~Sharp, K.~Goatman, G.~Prescott, G.~Scotland, A.~Fleming, S.~Philip, C.~Santiago, S.~Borooah, D.~Broadbent, V.~Chong, P.~Dodson, S.~Harding, G.~Leese, C.~Styles, K.~Swa, H.~Wharton, Improving the economic value of photographic screening for optical coherence tomography-detectable macular oedema: a prospective, multicentre, uk study, Health Technol Assess 17~(51) (2013) 1--142.
\newblock \href {https://doi.org/10.3310/hta17510} {\path{doi:10.3310/hta17510}}.

\bibitem{RN7}
N.~K. Waheed, J.~S. Duker, \href{https://doi.org/10.1007/s40135-013-0019-z}{Oct in the management of diabetic macular edema}, Current Ophthalmology Reports 1~(3) (2013) 128--133.
\newblock \href {https://doi.org/10.1007/s40135-013-0019-z} {\path{doi:10.1007/s40135-013-0019-z}}.
\newline\urlprefix\url{https://doi.org/10.1007/s40135-013-0019-z}

\bibitem{RN8}
D.~Koozekanani, C.~Roberts, S.~E. Katz, E.~E. Herderick, Intersession repeatability of macular thickness measurements with the humphrey 2000 oct, Invest Ophthalmol Vis Sci 41~(6) (2000) 1486--91.

\bibitem{RN9}
D.~A. Salz, A.~J. Witkin, Imaging in diabetic retinopathy, Middle East Afr J Ophthalmol 22~(2) (2015) 145--50.
\newblock \href {https://doi.org/10.4103/0974-9233.151887} {\path{doi:10.4103/0974-9233.151887}}.

\bibitem{transffa}
D.~Shi, W.~Zhang, S.~He, Y.~Chen, F.~Song, S.~Liu, R.~Wang, Y.~Zheng, M.~He, Translation of color fundus photography into fluorescein angiography using deep learning for enhanced diabetic retinopathy screening, Ophthalmology science 3~(4) (2023) 100401.

\bibitem{ffagpt}
X.~Chen, W.~Zhang, P.~Xu, Z.~Zhao, Y.~Zheng, D.~Shi, M.~He, Ffa-gpt: an automated pipeline for fundus fluorescein angiography interpretation and question-answer, NPJ digital medicine 7~(1) (2024) 111.

\bibitem{transicga}
R.~Chen, W.~Zhang, F.~Song, H.~Yu, D.~Cao, Y.~Zheng, M.~He, D.~Shi, Translating color fundus photography to indocyanine green angiography using deep-learning for age-related macular degeneration screening, NPJ digital medicine 7~(1) (2024) 34.

\bibitem{RN14}
J.~Sun, D.~Wei, L.~Wang, Y.~Zheng, Lesion guided explainable few weak-shot medical report generation, in: International Conference on Medical Image Computing and Computer-Assisted Intervention, 2022.

\bibitem{RN15}
F.~Tang, X.~Wang, A.~R. Ran, C.~K.~M. Chan, M.~Ho, W.~Yip, A.~L. Young, J.~Lok, S.~Szeto, J.~Chan, F.~Yip, R.~Wong, Z.~Tang, D.~Yang, D.~S. Ng, L.~J. Chen, M.~Brelén, V.~Chu, K.~Li, T.~H.~T. Lai, G.~S. Tan, D.~S.~W. Ting, H.~Huang, H.~Chen, J.~H. Ma, S.~Tang, T.~Leng, S.~Kakavand, S.~S. Mannil, R.~T. Chang, G.~Liew, B.~Gopinath, T.~Y.~Y. Lai, C.~P. Pang, P.~H. Scanlon, T.~Y. Wong, C.~C. Tham, H.~Chen, P.~A. Heng, C.~Y. Cheung, A multitask deep-learning system to classify diabetic macular edema for different optical coherence tomography devices: A multicenter analysis, Diabetes Care 44~(9) (2021) 2078--2088.
\newblock \href {https://doi.org/10.2337/dc20-3064} {\path{doi:10.2337/dc20-3064}}.

\bibitem{RN16}
E.~Midena, L.~Toto, L.~Frizziero, G.~Covello, T.~Torresin, G.~Midena, L.~Danieli, E.~Pilotto, M.~Figus, C.~Mariotti, M.~Lupidi, Validation of an automated artificial intelligence algorithm for the quantification of major oct parameters in diabetic macular edema, J Clin Med 12~(6) (2023).
\newblock \href {https://doi.org/10.3390/jcm12062134} {\path{doi:10.3390/jcm12062134}}.

\bibitem{RN17}
A.~Tripathi, P.~Kumar, V.~Mayya, A.~Tulsani, Generating oct b-scan dme images using optimized generative adversarial networks (gans), Heliyon 9~(8) (2023) e18773.
\newblock \href {https://doi.org/10.1016/j.heliyon.2023.e18773} {\path{doi:10.1016/j.heliyon.2023.e18773}}.

\bibitem{tang2017private}
F.~Y. Tang, D.~S. Ng, A.~Lam, F.~Luk, R.~Wong, C.~Chan, S.~Mohamed, A.~Fong, J.~Lok, T.~Tso, et~al., Determinants of quantitative optical coherence tomography angiography metrics in patients with diabetes, Scientific reports 7~(1) (2017) 2575.

\bibitem{srinivasan2014small}
P.~P. Srinivasan, L.~A. Kim, P.~S. Mettu, S.~W. Cousins, G.~M. Comer, J.~A. Izatt, S.~Farsiu, Fully automated detection of diabetic macular edema and dry age-related macular degeneration from optical coherence tomography images, Biomedical optics express 5~(10) (2014) 3568--3577.

\bibitem{niemeijer2009retinopathy}
M.~Niemeijer, B.~Van~Ginneken, M.~J. Cree, A.~Mizutani, G.~Quellec, C.~I. S{\'a}nchez, B.~Zhang, R.~Hornero, M.~Lamard, C.~Muramatsu, et~al., Retinopathy online challenge: automatic detection of microaneurysms in digital color fundus photographs, IEEE transactions on medical imaging 29~(1) (2009) 185--195.

\bibitem{diabetic-retinopathy-detection}
E.~Dugas, Jared, Jorge, W.~Cukierski, Diabetic retinopathy detection, \url{https://kaggle.com/competitions/diabetic-retinopathy-detection}, kaggle (2015).

\bibitem{kobat2022automated}
S.~G. Kobat, N.~Baygin, E.~Yusufoglu, M.~Baygin, P.~D. Barua, S.~Dogan, O.~Yaman, U.~Celiker, H.~Yildirim, R.-S. Tan, et~al., Automated diabetic retinopathy detection using horizontal and vertical patch division-based pre-trained densenet with digital fundus images, Diagnostics 12~(8) (2022) 1975.

\bibitem{porwal2020idrid}
P.~Porwal, S.~Pachade, M.~Kokare, G.~Deshmukh, J.~Son, W.~Bae, L.~Liu, J.~Wang, X.~Liu, L.~Gao, et~al., Idrid: Diabetic retinopathy--segmentation and grading challenge, Medical image analysis 59 (2020) 101561.

\bibitem{liu2022deepdrid}
R.~Liu, X.~Wang, Q.~Wu, L.~Dai, X.~Fang, T.~Yan, J.~Son, S.~Tang, J.~Li, Z.~Gao, et~al., Deepdrid: Diabetic retinopathy—grading and image quality estimation challenge, Patterns 3~(6) (2022).

\bibitem{qian2024drac}
B.~Qian, H.~Chen, X.~Wang, Z.~Guan, T.~Li, Y.~Jin, Y.~Wu, Y.~Wen, H.~Che, G.~Kwon, et~al., Drac 2022: A public benchmark for diabetic retinopathy analysis on ultra-wide optical coherence tomography angiography images, Patterns 5~(3) (2024).

\bibitem{orlando2020refuge}
J.~I. Orlando, H.~Fu, J.~B. Breda, K.~Van~Keer, D.~R. Bathula, A.~Diaz-Pinto, R.~Fang, P.-A. Heng, J.~Kim, J.~Lee, et~al., Refuge challenge: A unified framework for evaluating automated methods for glaucoma assessment from fundus photographs, Medical image analysis 59 (2020) 101570.

\bibitem{fang2022refuge2}
H.~Fang, F.~Li, J.~Wu, H.~Fu, X.~Sun, J.~Son, S.~Yu, M.~Zhang, C.~Yuan, C.~Bian, et~al., Refuge2 challenge: A treasure trove for multi-dimension analysis and evaluation in glaucoma screening, arXiv preprint arXiv:2202.08994 (2022).

\bibitem{de2023airogs}
C.~De~Vente, K.~A. Vermeer, N.~Jaccard, H.~Wang, H.~Sun, F.~Khader, D.~Truhn, T.~Aimyshev, Y.~Zhanibekuly, T.-D. Le, et~al., Airogs: Artificial intelligence for robust glaucoma screening challenge, IEEE transactions on medical imaging 43~(1) (2023) 542--557.

\bibitem{lemij2023characteristics}
H.~G. Lemij, C.~de~Vente, C.~I. S{\'a}nchez, K.~A. Vermeer, Characteristics of a large, labeled data set for the training of artificial intelligence for glaucoma screening with fundus photographs, Ophthalmology Science 3~(3) (2023) 100300.

\bibitem{ODIR-2019}
Peking university international competition on ocular disease intelligent recognition (odir-2019), URL:\url{https://odir2019.grand-challenge.org/} (2019).

\bibitem{pachade2025rfmid}
S.~Pachade, P.~Porwal, M.~Kokare, G.~Deshmukh, V.~Sahasrabuddhe, Z.~Luo, F.~Han, Z.~Sun, L.~Qihan, S.-i. Kamata, et~al., Rfmid: Retinal image analysis for multi-disease detection challenge, Medical Image Analysis 99 (2025) 103365.

\bibitem{fang2024open}
H.~Fang, F.~Li, J.~Wu, H.~Fu, X.~Sun, J.~I. Orlando, H.~Bogunovi{\'c}, X.~Zhang, Y.~Xu, Open fundus photograph dataset with pathologic myopia recognition and anatomical structure annotation, Scientific Data 11~(1) (2024) 99.

\bibitem{fang2022adam}
H.~Fang, F.~Li, H.~Fu, X.~Sun, X.~Cao, F.~Lin, J.~Son, S.~Kim, G.~Quellec, S.~Matta, et~al., Adam challenge: Detecting age-related macular degeneration from fundus images, IEEE transactions on medical imaging 41~(10) (2022) 2828--2847.

\bibitem{fu2020age}
H.~Fu, F.~Li, X.~Sun, X.~Cao, J.~Liao, J.~I. Orlando, X.~Tao, Y.~Li, S.~Zhang, M.~Tan, et~al., Age challenge: angle closure glaucoma evaluation in anterior segment optical coherence tomography, Medical Image Analysis 66 (2020) 101798.

\bibitem{zhang2023angiographic}
W.~Zhang, P.~Chotcomwongse, X.~Chen, F.~H. Chung, F.~Song, X.~Zhang, M.~He, D.~Shi, P.~Ruamviboonsuk, Angiographic report generation for the 3rd aptos’s competition: Dataset and baseline methods, medRxiv (2023) 2023--11.

\bibitem{bogunovic2019retouch}
H.~Bogunovi{\'c}, F.~Venhuizen, S.~Klimscha, S.~Apostolopoulos, A.~Bab-Hadiashar, U.~Bagci, M.~F. Beg, L.~Bekalo, Q.~Chen, C.~Ciller, et~al., Retouch: The retinal oct fluid detection and segmentation benchmark and challenge, IEEE transactions on medical imaging 38~(8) (2019) 1858--1874.

\bibitem{ROCC-2017}
R.~Rasti, H.~Rabbani, Retinal oct classification challenge (rocc), URL:\url{https://rocc.grand-challenge.org/} (2017).

\bibitem{wu2023gamma}
J.~Wu, H.~Fang, F.~Li, H.~Fu, F.~Lin, J.~Li, Y.~Huang, Q.~Yu, S.~Song, X.~Xu, et~al., Gamma challenge: glaucoma grading from multi-modality images, Medical Image Analysis 90 (2023) 102938.

\bibitem{APTOS-2024}
2024 aptos big data competition, URL:\url{https://2024.asiateleophth.org/big-data-competition/} (2024).

\bibitem{RN34}
M.~H. Akpinar, A.~Sengur, O.~Faust, L.~Tong, F.~Molinari, U.~R. Acharya, Artificial intelligence in retinal screening using oct images: A review of the last decade (2013–2023), Computer Methods and Programs in Biomedicine 254 (2024) 108253.
\newblock \href {https://doi.org/https://doi.org/10.1016/j.cmpb.2024.108253} {\path{doi:https://doi.org/10.1016/j.cmpb.2024.108253}}.

\bibitem{RN51}
T.~Sonobe, H.~Tabuchi, H.~Ohsugi, H.~Masumoto, N.~Ishitobi, S.~Morita, H.~Enno, D.~Nagasato, Comparison between support vector machine and deep learning, machine-learning technologies for detecting epiretinal membrane using 3d-oct, International Ophthalmology 39~(8) (2019) 1871--1877.
\newblock \href {https://doi.org/10.1007/s10792-018-1016-x} {\path{doi:10.1007/s10792-018-1016-x}}.

\bibitem{RN50}
A.~Mihalache, R.~S. Huang, M.~M. Popovic, N.~S. Patil, B.~U. Pandya, R.~Shor, A.~Pereira, J.~M. Kwok, P.~Yan, D.~T. Wong, P.~J. Kertes, R.~H. Muni, Accuracy of an artificial intelligence chatbot’s interpretation of clinical ophthalmic images, JAMA Ophthalmology 142~(4) (2024) 321--326.
\newblock \href {https://doi.org/10.1001/jamaophthalmol.2024.0017} {\path{doi:10.1001/jamaophthalmol.2024.0017}}.

\bibitem{RN52}
M.~A. Hussain, A.~Bhuiyan, C.~D.~Luu, R.~Theodore~Smith, R.~H.~Guymer, H.~Ishikawa, J.~S.~Schuman, K.~Ramamohanarao, Classification of healthy and diseased retina using sd-oct imaging and random forest algorithm, PLOS ONE 13~(6) (2018) 1--17.
\newblock \href {https://doi.org/10.1371/journal.pone.0198281} {\path{doi:10.1371/journal.pone.0198281}}.

\bibitem{RN53}
H.-Y. Chen, M.-L. Huang, P.-T. Hung, Logistic regression analysis for glaucoma diagnosis using stratus optical coherence tomography, Optometry and Vision Science 83~(7) (2006) 00023.
\newblock \href {https://doi.org/10.1097/00006324-200607000-00023} {\path{doi:10.1097/00006324-200607000-00023}}.

\bibitem{RN54}
P.~Seeb{\"o}ck, S.~M. Waldstein, S.~Klimscha, H.~Bogunovic, T.~Schlegl, B.~S. Gerendas, R.~Donner, U.~Schmidt-Erfurth, G.~Langs, Unsupervised identification of disease marker candidates in retinal oct imaging data, IEEE transactions on medical imaging 38~(4) (2018) 1037--1047.

\bibitem{RN55}
M.~Awais, H.~M{\"u}ller, T.~B. Tang, F.~Meriaudeau, Classification of sd-oct images using a deep learning approach, in: 2017 IEEE International Conference on Signal and Image Processing Applications (ICSIPA), IEEE, 2017, pp. 489--492.

\bibitem{RN56}
D.~Feng, X.~Chen, Z.~Zhou, H.~Liu, Y.~Wang, L.~Bai, S.~Zhang, X.~Mou, A preliminary study of predicting effectiveness of anti-vegf injection using oct images based on deep learning, in: 2020 42nd annual international conference of the IEEE engineering in medicine \& biology society (EMBC), IEEE, 2020, pp. 5428--5431.

\bibitem{RN57}
J.~Wang, G.~Deng, W.~Li, Y.~Chen, F.~Gao, H.~Liu, Y.~He, G.~Shi, Deep learning for quality assessment of retinal oct images, Biomedical optics express 10~(12) (2019) 6057--6072.

\bibitem{RN58}
R.~M. Kamble, G.~C. Chan, O.~Perdomo, M.~Kokare, F.~A. Gonzalez, H.~M{\"u}ller, F.~M{\'e}riaudeau, Automated diabetic macular edema (dme) analysis using fine tuning with inception-resnet-v2 on oct images, in: 2018 IEEE-EMBS Conference on Biomedical Engineering and Sciences (IECBES), IEEE, 2018, pp. 442--446.

\bibitem{RN59}
Y.~Shen, J.~Li, W.~Zhu, K.~Yu, M.~Wang, Y.~Peng, Y.~Zhou, L.~Guan, X.~Chen, Graph attention u-net for retinal layer surface detection and choroid neovascularization segmentation in oct images, IEEE Transactions on Medical Imaging 42~(11) (2023) 3140--3154.

\bibitem{RN60}
N.~Ndipenoch, A.~Miron, Y.~Li, Performance evaluation of retinal oct fluid segmentation, detection, and generalization over variations of data sources, IEEE Access 12 (2024) 31719--31735.

\bibitem{shi2024eyefound}
D.~Shi, W.~Zhang, X.~Chen, Y.~Liu, J.~Yang, S.~Huang, Y.~C. Tham, Y.~Zheng, M.~He, Eyefound: a multimodal generalist foundation model for ophthalmic imaging, arXiv preprint arXiv:2405.11338 (2024).

\bibitem{zhang2024fundus2video}
W.~Zhang, S.~Huang, J.~Yang, R.~Chen, Z.~Ge, Y.~Zheng, D.~Shi, M.~He, Fundus2video: Cross-modal angiography video generation from static fundus photography with clinical knowledge guidance, in: International Conference on Medical Image Computing and Computer-Assisted Intervention, Springer, 2024, pp. 689--699.

\bibitem{RN61}
D.~Shi, W.~Zhang, J.~Yang, S.~Huang, X.~Chen, M.~Yusufu, K.~Jin, S.~Lin, S.~Liu, Q.~Zhang, et~al., Eyeclip: A visual-language foundation model for multi-modal ophthalmic image analysis, arXiv preprint arXiv:2409.06644 (2024).

\bibitem{RN62}
P.~Xu, X.~Chen, Z.~Zhao, D.~Shi, Unveiling the clinical incapabilities: a benchmarking study of gpt-4v (ision) for ophthalmic multimodal image analysis, British Journal of Ophthalmology 108~(10) (2024) 1384--1389.

\bibitem{RN63}
F.~Cappellani, K.~R. Card, C.~L. Shields, J.~S. Pulido, J.~A. Haller, Reliability and accuracy of artificial intelligence chatgpt in providing information on ophthalmic diseases and management to patients, Eye 38~(7) (2024) 1368--1373.

\bibitem{RN18}
Z.~Liu, Y.~Lin, Y.~Cao, H.~Hu, Y.~Wei, Z.~Zhang, S.~Lin, B.~Guo, Swin transformer: Hierarchical vision transformer using shifted windows, in: Proceedings of the IEEE/CVF international conference on computer vision, 2021, pp. 10012--10022.

\bibitem{RN19}
M.-A. Carbonneau, V.~Cheplygina, E.~Granger, G.~Gagnon, Multiple instance learning: A survey of problem characteristics and applications, Pattern Recognition 77 (2018) 329--353.

\bibitem{RN20}
K.~He, X.~Zhang, S.~Ren, J.~Sun, Deep residual learning for image recognition, in: Proceedings of the IEEE conference on computer vision and pattern recognition, 2016, pp. 770--778.

\bibitem{RN21}
G.~K. Wallace, The jpeg still picture compression standard, IEEE transactions on consumer electronics 38~(1) (1992) xviii--xxxiv.

\bibitem{RN22}
W.~Sultani, C.~Chen, M.~Shah, Real-world anomaly detection in surveillance videos, in: Proceedings of the IEEE conference on computer vision and pattern recognition, 2018, pp. 6479--6488.

\bibitem{RN23}
D.~Shanmugam, D.~Blalock, G.~Balakrishnan, J.~Guttag, Better aggregation in test-time augmentation, in: Proceedings of the IEEE/CVF international conference on computer vision, 2021, pp. 1214--1223.

\bibitem{RN24}
M.~Tan, Q.~Le, Efficientnet: Rethinking model scaling for convolutional neural networks, in: International conference on machine learning, PMLR, 2019, pp. 6105--6114.

\bibitem{RN25}
M.~Sandler, A.~Howard, M.~Zhu, A.~Zhmoginov, L.-C. Chen, Mobilenetv2: Inverted residuals and linear bottlenecks, in: Proceedings of the IEEE conference on computer vision and pattern recognition, 2018, pp. 4510--4520.

\bibitem{RN26}
J.~Deng, W.~Dong, R.~Socher, L.-J. Li, K.~Li, L.~Fei-Fei, Imagenet: A large-scale hierarchical image database, in: 2009 IEEE conference on computer vision and pattern recognition, Ieee, 2009, pp. 248--255.

\bibitem{RN27}
G.~Huang, Z.~Liu, L.~Van Der~Maaten, K.~Q. Weinberger, Densely connected convolutional networks, in: Proceedings of the IEEE conference on computer vision and pattern recognition, 2017, pp. 4700--4708.

\bibitem{RN28}
J.~H. Friedman, Greedy function approximation: a gradient boosting machine, Annals of statistics (2001) 1189--1232.

\bibitem{RN29}
M.~R. Segal, Machine learning benchmarks and random forest regression (2004).

\bibitem{RN30}
G.~Panozzo, E.~Gusson, B.~Parolini, A.~Mercanti, Role of oct in the diagnosis and follow up of diabetic macular edema, in: Seminars in ophthalmology, Vol.~18, Taylor \& Francis, 2003, pp. 74--81.

\bibitem{RN31}
X.~Wang, F.~Tang, H.~Chen, C.~Y. Cheung, P.-A. Heng, Deep semi-supervised multiple instance learning with self-correction for dme classification from oct images, Medical Image Analysis 83 (2023) 102673.

\bibitem{RN32}
N.~Padmasini, R.~Umamaheswari, M.~Y. Sikkandar, M.~D. Sindal, Comparison of deep CNNs in the identification of DME structural changes in retinal OCT scans, Elsevier, 2023, pp. 35--51.

\bibitem{RN33}
Z.~Qi, S.~Khorram, F.~Li, Visualizing deep networks by optimizing with integrated gradients, in: CVPR Workshops, Vol.~2, 2019, pp. 1--4.

\end{thebibliography}

\pagebreak
\appendix
    \section{APTOS 2021 Team Introduction}
\textbf{BlueSky:} \textbf{1.} Ruijie Yao (ruijie.yao@duke.edu) is a first-year Ph.D. student in the Mechanical Engineering and Materials Science (MEMS) program at Duke University. He received his master’s degree in 2024 and his bachelor’s degree in 2021, both from the Department of Electronic Engineering at Tsinghua University. His research interests lie in the areas of computer vision, machine learning, and medical intelligence. \textbf{2.} Lianhao Zhou (leozhou@tamu.edu) is a first-year Ph.D. student in Computer Science at Texas A\&M University. He received both his Bachelor's and Master's degrees from the Department of Electronic Engineering at Tsinghua University. His research interests include AI for Science and LLM Agent. 
\textbf{3.} Yuxuan Zhou (zhou-yx21@mails.tsinghua.edu.cn) is a fourth-year Ph.D. student at the department of Electronic Engineering, Tsinghua University. He also received his Bachelor's degree from the department of Electronic Engineering at Tsinghua University. His research interests include LLM medical evaluation and domain-specific training.

\textbf{LightRain:}
Hui Feng(fenghuiwk@gmail.com) is from Xidian University. His research interests are computer vision, video anomaly detection, multimodal generation, and applications of large language models.

\textbf{DarkStyle:}
DarkStyle team consists of Qiping Zhou (qiping.zhou@intalight.com), an Algorithm Engineer at Svision Imaging Ltd., specializing in deep learning solutions for medical image analysis.
\label{app1}


\typeout{Generating .bbl file...}
\end{document}